\documentclass[12pt,twoside]{article}
\usepackage{amsmath,amssymb}
\def\1#1{{\bf #1}}
\def\2#1{{\cal #1}}\def\9#1{{\sl #1}}\def\4#1{{\tt #1}}\def\5#1{{\sf #1}}
\def\6#1{{\mathfrak #1}}\def\7#1{{\mathbb #1}}\def\8#1{{\rm #1}}
\def\9#1{{\mit #1}}

\def\aut{{\rm Aut}}
\def\3{{\ss}}

\def\ol{\overline}

\def\skb{\vskip 0.5cm}

\def\beq{\begin{eqnarray}}
\def\eeq{\end{eqnarray}}
\def\vs{\vspace{0.2cm} \\}

\newtheorem{The}{Theorem}[section]
\newtheorem{Def}[The]{Definiton}
\newtheorem{Lem}[The]{Lemma}
\newtheorem{Pro}[The]{Proposition}
\newtheorem{Cor}[The]{Corollary}
\def\bdef{\begin{Def}\1: \em}
\def\eef{\end{Def}}
\def\bnoa{\paragraph*{\it Notation:}}
\def\enoa{\skb}
\def\blem{\begin{Lem}\1: }
\def\elem{\end{Lem}}
\def\bthe{\begin{The}\1: }
\def\ethe{\end{The}}
\def\bpro{\begin{Pro}\1: }
\def\epro{\end{Pro}}
\def\bcor{\begin{Cor}\1: }
\def\ecor{\end{Cor}}
\def\bobs{\paragraph*{\it Observation:}}
\def\eobs{\skb}
\def\brem{\paragraph*{\it Remark:}}
\def\erem{\skb}
\def\supp{{\rm supp}}

\def\ps{{\rm  ps}}

\def\id{{\rm id}}

\def\al{\alpha}
\def\be{\beta}
\def\gam{\gamma}\def\Gam{\Gamma}
\def\lam{\lambda}\def\Lam{\Lambda}
\def\eps{\epsilon} 
\def\te{\theta}

\def\Sgm{\Sigma}
\def\om{\omega}\def\Om{\Omega}

\def\bpr{\paragraph*{\it Proof.}}

\def\epr{$\square$\skb}

\def\pa{\partial}
\def\<{\langle}
\def\>{\rangle}
\def\Ad{{\rm Ad}}

\def\dyn{{\rm dyn}}

\def\bl{\biggl}
\def\br{\biggr}

\def\bs{\backslash}

\def\bdes{\begin{enumerate}}
\def\edes{\end{enumerate}}
\newcommand\itno[1]{\item[{\it ({#1})}]}

\def\bmat{\left( \begin{array}{ccc} }
\def\emat{\end{array} \right)}

\def\dimker{{\rm dim \ ker}}
\def\dimcoker{{\rm dim \ coker}}
\def\ind{{\rm ind}}
\def\beqa{\begin{eqnarray*}}
\def\eeqa{\end{eqnarray*}}
\def\bdia{\begin{diagram}}
\def\edia{\end{diagram}}

\def\olt{\ \overline{\otimes}\ }

\def\deg{{\rm deg}}

\title{\bf
Construction of Kink Sectors for Two-Dimensional 
Quantum Field Theory Models \\ (An Algebraic Approach)}
\author{{\it Dirk Schlingemann} \\
II. Institut f\"ur Theoretische Physik \\
Universit\"at Hamburg \\ and \\
Erwin Schr\"odinger International Institute \\ 
for Mathematical Physics (ESI)\\
Vienna}
\begin{document}
\maketitle
\abstract{
Several two-dimensional quantum field theory models 
have more than one vacuum state. Familiar examples are
the Sine-Gordon and the $\phi^4_2$-model. It is known 
that in these models there are also states, 
called kink states, which interpolate different vacua. 
A general construction scheme for 
kink states in the framework of algebraic quantum field theory is 
developed in a previous paper.
However, for the application of this method, the crucial condition is the  
split property for wedge algebras 
in the vacuum representations of the considered models. 
It is believed that the vacuum representations of 
$P(\phi)_2$-models fulfill this condition, but a 
rigorous proof is only known for the massive free scalar field. 
Therefore, we investigate in a  
construction of kink states which can directly be 
applied to a large class of quantum field theory models, by making use of 
the properties of the dynamics of a $P(\phi)_2$ and Yukawa$_2$ models.}
\thispagestyle{empty}
\newpage
\section{Introduction}
\label{s1}

Studying $1+1$-dimensional quantum field theories from an 
axiomatic point of view shows that kink sectors naturally appear in the 
theory of superselection sectors \cite{Fre1,Fre2,Schl2}. 
This paper is concerned with the 
construction of kink sectors for concrete quantum field theory models, like 
$P(\phi)_2$ and Yukawa$_2$ models.\footnote{Parts are extracted from
the PhD thesis \cite{Schl6}.}  

Our subsequent analysis is placed into the framework of 
algebraic quantum field theory which has 
turned out to be a successful 
formalism to describe physical concepts like observables, states , 
superselection sectors (charges) and statistics.
These notions can appropriately be described by mathematical concepts
like C*-algebras, positive linear functionals and equivalence classes 
of representations. For the convenience of the reader, 
we shall state the relevant 
definitions and assumptions here.

Let $\2O\subset \7R^{1,s}$ be a region in space-time. We denote by 
$\6A(\2O)$ the algebra generated by all observables which can be measured 
within $\2O$. For technical reasons we always suppose that 
$\6A(\2O)$ is a C*-algebra and $\2O$ is a double cone, 
i.e. a bounded and causally complete region.
Motivated by physical principles, we make the following assumptions:
 
\bdes
\itno 1
The assignment 
\beqa
\6A:\2O\mapsto\6A(\2O)
\eeqa
is an isotonous net of C*-algebras, i.e.  
if $\2O_1$ is contained in $\2O_2$, then $\6A(\2O_1)$ is a C*-sub-algebra 
of $\6A(\2O_2)$. The isotony encodes the fact that 
each observable which can be measured within $\2O$ 
can also be measured in every larger region.
Furthermore, the C*-inductive limit
\beqa
C^*(\6A)
\eeqa
of the net $\6A$
can be constructed since the set of double cones is directed. 
We refer to \cite{Sak} for this notion. 
\itno 2
Two local operations which take place 
in space-like separated regions should not influence each other.
The {\em principle of locality} is formulated as follows:
If the regions $\2O_1$ and $\2O_2$ are space-like separated, 
then the elements of $\6A(\2O_1)$ commute with those of $\6A(\2O_2)$. 
\itno 3
Each operator $a$ which is localized in a region $\2O$ should have an 
equivalent counterpart which is localized in the translated region
$\2O+x$. The {\em principle of translation symmetry} is encoded by 
the existence of an automorphism group $\{\al_x;x\in\7R^{1,s}\}$ which 
acts on the C*-algebra $C^*(\6A)$ such that 
$\al_x$ maps $\6A(\2O)$ onto $\6A(\2O+x)$.
\edes

A net of C*-algebras which fulfills the conditions 
{\it (1)} to {\it (3)} is called a 
{\em translationally covariant Haag-Kastler net}.

In order to discuss particle-like concepts, we  
select an appropriate class $\6S$ of normalized 
positive linear functionals, called states, of $C^*(\6A)$. 
We require that the states $\om\in\6S$ fulfill the conditions:
\bdes
\itno 1 
There exists a strongly continuous unitary representation
of the translation group $U:x\mapsto U(x)$ on the GNS\footnote{
Given a state $\om\in\6S$, we obtain 
via GNS-construction a Hilbert space $\2H$, 
a *-representation $\pi$ of $C^*(\6A)$ 
on $\2H$ and a vector $\Om\in\2H$ such that 
$\<\Om,\pi(a)\Om\>=\om(a)$ for each $a\in C^*(\6A)$. 
The triple $(\2H,\pi,\Om)$ is called the GNS-triple of $\om$.}-Hilbert 
space $\2H$ 
which implements the translations in the GNS-representation $\pi$, i.e.
\beqa
\pi(\al_xa)=U(x)\pi(a)U(-x)
\eeqa 
for each $a\in C^*(\6A)$.
\itno 2 
The stability of a physical system is encoded in the 
spectrum condition (positivity of the energy), i.e. 
the spectrum (of the generator) of $U(x)$ is contained 
in the closed forward light cone.
\edes
These conditions are also known as the {\em Borchers criterion}. 
States which satisfy the Borchers criterion 
and which are, in addition, translationally invariant are 
called {\em vacuum states}.

Kinks already appear in classical 
field theories and the typical systems in which they occur are 
1+1-dimensional. Familiar examples are
the Sine-Gordon and the $\phi^4_2$-model. 
We briefly describe the latter:

The Lagrangian density of the model is given by 
\beqa
\6L(\phi,x)={1\over 2} \ \pa_\mu\phi(x)\pa^\mu\phi(x)-U(\phi(x))
\eeqa
where the potential $U$ is given by
\beqa
U(z):=\lam/2 \ (z^2-a)^2 \ \ .
\eeqa
The energy of a classical field configuration $\phi$ is  
\beqa
E(\phi)=\int \8d\1x \ \biggl ( {1\over 2} (\pa_0\phi(0,\1x))^2  
+  {1\over 2} (\pa_1\phi(0,\1x))^2
+  U(\phi(0,\1x)) \biggr )\ \ .
\eeqa
With the choice of $U$, given above, the absolute minimum value of $U$ is 
zero and thus the energy functional $E:\phi\mapsto E(\phi)$ is positive. 

There are two configurations $\phi_\pm$ with 
zero energy $E(\phi_\pm)=0$: 
\beqa
\phi_\pm:(t,\1x)\mapsto \pm \ a \ \ .
\eeqa
These configurations are invariant under space-time translations and 
represent the vacua of the classical system. 

There are two further configurations $\phi_s,\phi_{\bar s}$ which 
are stationary points of the energy functional $E$.
They are given by 
\beqa
\phi_s:(t,\1x)\mapsto a \ \tanh(\ \sqrt{\lam} a \1x \ )
\ \ \mbox{ and } \ \ 
\phi_{\bar s}:(t,\1x)\mapsto -a \ \tanh(\ \sqrt{\lam} a \1x \ ) \ \ .
\eeqa
These configurations represent the kinks of the classical system which 
{\em interpolate} 
the vacua $\phi_\pm$. Indeed, we have for the kink $\phi_s$
\beq\label{classinterpol}
\lim_{\1x\to\pm\infty}\phi_s(t,\1x)=\phi_\pm(t,\1x)=\pm a \ \ .
\eeq
The configuration $\phi_{\bar s}$, 
which interpolates the vacua $\phi_\pm$ in the opposite direction,  
represents the anti-kink of $\phi_s$. 
Both of them have the same energy, namely
\beqa
E(\phi_s)=E(\phi_{\bar s})={4\over 3} \sqrt{\lam} a^3 \ \ .
\eeqa

From the classical example above, we see that 
the crucial properties of a kink are to interpolate vacuum configurations 
as well as to be a configuration of finite energy. 
Motivated by these properties, in quantum field theory 
a {\em kink state} $\om$ is defined as 
follows: 

\paragraph*{\it The interpolation property:}
For each observable $a$, the limits 
\beq\label{qminterpol}
\lim_{\1x\to\pm\infty}\om(\al_{(t,\1x)}(a))=\om_\pm(a)
\eeq
exist and $\om_\pm$ are vacuum states. Note that equation (\ref{qminterpol})
is the quantum version of the 
interpolation property (\ref{classinterpol}).

\paragraph*{\it Positivity of the energy:}
$\om$ fulfills the Borchers criterion. 
\skb

In the literature the concept of {\em kink} as
described above is often called {\em soliton} (see \cite{Froh1,Froh3}) 
or more seldom {\em lump} (see \cite{Col}).
In the following, we shall use the word kink.   

In \cite{Schl4}, 
a construction scheme for kink states has been developed which is based on 
general principles. In order to make the comprehension of the subsequent   
sections easier we shall state the main ideas here.  
The construction of an interpolating kink state is based on a 
simple physical idea: 
Let $\6A$ be a Haag-Kastler net of W*-algebras in $1+1$-dimensions.
Each double cone $\2O$ splits our system into two infinitely
extended laboratories, namely the laboratory which belongs to the left 
space-like complement $\2O_{LL}$, and the laboratory 
$\2O_{RR}$ which belongs the right space-like complement $\2O_{RR}$. 
In order to prepare an interpolating kink state, we wish to 
prepare one vacuum state $\om_1$ in the left laboratory 
$\2O_{LL}$, and another vacuum state $\om_2$ in the 
right laboratory $\2O_{RR}$. 
This can only be done if the 
preparation of $\om_1$ does not disturb the preparation procedure 
of $\om_2$. In other words, 
the physical operations which take place in 
the laboratory on the left side $\2O_{LL}$ should be statistically 
independent of those which take place in $\2O_{RR}$.

Therefore, we require that there exists a 
vacuum representation $\pi_0$ such that the W*-tensor product  
\beqa
\6A_{\pi_0}(\2O_{LL})\olt\6A_{\pi_0}(\2O_{RR})
\eeqa
is unitarily isomorphic to the von Neumann algebra
\beqa
\6A_{\pi_0}(\2O_{LL})\vee\6A_{\pi_0}(\2O_{RR})
\eeqa
where $\6A_{\pi_0}$ is the net in the vacuum representation $\pi_0$.
\footnote{ 
for an unbounded region $\2U$, $\6A_{\pi_0}(\2U)$ denotes the 
von Neumann algebra which is generated by all local algebras 
$\6A_{\pi_0}(\2O)$ with $\2O\subset\2U$.} 
This condition is equivalent to the existence of a type I factor $\2N$ which 
sits between $\6A_{\pi_0}(\2O_{RR})$ and $\6A_{\pi_0}(\2O_R)$:
\beqa
\6A_{\pi_0}(\2O_{RR})\subset\2N\subset\6A_{\pi_0}(\2O_R) \ \ .
\eeqa
Here $\2O_R$ is the space-like complement of $\2O_{LL}$. 
In other words, the inclusion 
\beq\label{split1}
\6A_{\pi_0}(\2O_{RR})\subset\6A_{\pi_0}(\2O_R)
\eeq 
is {\em split}.  

A detailed investigation of standard split inclusions of W*-algebras 
has been carried out by S. Doplicher and R. Longo \cite{DLo}. 
We also refer to the results of D. Buchholz \cite{Bu1}, 
C. D'Antoni and R. Longo \cite{AntLo} and 
C. D'Antoni and K. Fredenhagen \cite{AntFre}. 

Let $\om_1$ and $\om_2$ be two 
inequivalent vacuum states whose restrictions 
to each local algebra $\6A(\2O)$ are normal. 
Using the isomorphy 
\beqa
\6A_{\pi_0}(\2O_{LL})\olt\6A_{\pi_0}(\2O_{RR})
\cong 
\6A_{\pi_0}(\2O_{LL})\vee\6A_{\pi_0}(\2O_{RR})
\eeqa
we conclude that the map
\beqa
ab\mapsto \om_1(a)\om_2(b) \ \ \mbox{, $a$ is localized in $\2O_{LL}$ 
and $b$ is localized in $\2O_{RR}$,}
\eeqa
defines a state of the algebra $C^*(\6A,\2O_{LL}\cup\2O_{RR})$ which, 
by the Hahn-Banach theorem, can be extended to a state $\om$ of the C*-algebra
of all observables.
The state $\om$ interpolates the vacua 
$\om_1$ and $\om_2$ correctly, but
for an explicit construction 
of an interpolating state which satisfies the {\em Borchers criterion}, 
some technical difficulties have to be overcome.

The condition that the inclusion (\ref{split1}) is split is 
sufficient to develop a general construction scheme for 
interpolating kink states. 
We shall give a brief description of it here. 

\paragraph*{\it Step 1:}
We consider the W*-tensor product of the net $\6A$ with itself:
\beqa
\6A\olt\6A:\2O\mapsto\6A(\2O)\olt\6A(\2O)
\eeqa
The map $\al_F$ which is given by interchanging the
tensor factors, 
\beqa
\al_F:a_1\otimes a_2\mapsto a_2\otimes a_1
\eeqa
is called the {\em flip automorphism}.
Since the inclusion (\ref{split1}) is split, the flip automorphism 
is unitarily implemented on $\6A_{\pi_0}\olt\6A_{\pi_0}(\2O_{RR})$ 
by a unitary operator $\te$
which is contained in $\6A_{\pi_0}\olt\6A_{\pi_0}(\2O_R)$ \cite{AntLo}.
The adjoint action of $\te$ induces an automorphism 
\beqa
\be:=(\pi_0\otimes\pi_0)^{-1}\circ\Ad(\te)\circ\pi_0\otimes\pi_0
\eeqa
which maps local algebras into local algebras. 
Here we have assumed that the representation $\pi_0$ is faithful in order
to build the inverse $\pi_0^{-1}$. For each 
observable $a$ which is localized in the left space-like complement of
$\2O$ we have $\be(a)=a$, and for each  
observable $b$ which is localized in the right space-like complement of
$\2O$ we have $\be(b)=\al_F(b)$.
Note that $\be$ may depend on the choice of the 
vacuum representation $\pi_0$.

\paragraph*{\it Step 2:}
It is obvious that the state 
\beqa
\om:=\om_1\otimes\om_2\circ\be|_{C^*(\6A)\otimes\11}
\eeqa
interpolates $\om_1$ and $\om_2$. 
Let $\pi_1$ and $\pi_2$ be the GNS-representations of $\om_1$ and $\om_2$ 
respectively. Then the GNS-representation 
\beqa
\pi=\pi_1\otimes\pi_2\circ\be|_{C^*(\6A\otimes\11)}
\eeqa
of $\om$ is translationally covariant because the automorphism 
\beqa
\al_x\circ\be\circ\al_{-x}\circ\be
\eeqa
is implemented by a cocycle $\gam(x)$ of {\em local} operators in 
$C^*(\6A)$. The positivity of the energy can be proven by showing the 
additivity of the energy-momentum spectrum for automorphisms like $\be$.  
This together implies that $\om$ is an interpolating kink state. 
\skb

In comparison to the work of J. Fr\"ohlich in which 
the existence of kink states for the $\phi^4_2$ and the 
Sine-Gordon model is proven \cite{Froh1,Froh3}, our
construction scheme has the following advantages:
\bdes
\item[$\oplus$]
It is independent of specific details of the considered model because 
the split property (\ref{split1}), which is the crucial condition 
for applying the construction scheme, can be motivated by general principles.
\item[$\oplus$]
It can be applied to pairs of vacuum sectors which are not  
related by a symmetry transformation, whereas the techniques of J. Fr\"ohlich 
rely on the existence of a symmetry transformation connecting different vacua.
Indeed, according to J. Z. Imbrie
\cite{Imb81}, there are examples for $P(\phi)_2$ models possessing 
more than one vacuum state, but where the different vacua are 
{\em not} related by a symmetry. We also mention here 
the papers of K. Gawedzki \cite{Gaw78} and S.J. Summers \cite{Sum2}.  
\edes  

Unfortunately, there is one disadvantage which is the price
we have to pay for using a model independent analysis.  
\bdes
\item[$\ominus$]
The split property for wedge algebras (\ref{split1}) has to be proven 
for the vacuum states of the model 
under consideration if we want to apply our construction scheme to it. 
It is believed that the
vacuum states of the $P(\phi)_2$- and $\mbox{Yukawa}_2$ models 
fulfill this condition, but a 
rigorous proof is only known for the massive free Bose and Fermi field 
\cite{AntFre,Bu1,Sum1}. 
\edes

In the present paper, we investigate 
an alternative construction of kink states which can 
directly be applied to models. 
It is convenient to formulate our setup in  
the time slice formulation of a quantum field theory. 
The time slice-formulation has two main aspects.
First, the {\em Cauchy data} with respect to a given space-like plane 
$\Sgm$ which describes the 
boundary conditions at time $t=0$. Second, the {\em dynamics} which 
describes the time evolution of the quantum fields. 

The {\em Cauchy data} of a quantum field theory are given by a 
net of v.Neumann-algebras
$$
\6M:=\{\6M(\2I)\subset\6B(\2H_0);
\mbox{ $\2I$ is open and bounded interval in $\Sgm$} \}
$$
represented on a Hilbert-space $\2H_0$. This net has to 
satisfy the following conditions:
\bdes
\itno 1
The net is isotonous, i.e. if  
$\2I_1\subset\2I_2$, then $\6M(\2I_1)\subset\6M(\2I_2)$.
\itno 2
The net is local, i.e. if 
$\2I_1\cap\2I_2=\emptyset$, then $\6M(\2I_1)\subset\6M(\2I_2)'$.
\itno 3
There exists a unitary and strongly continuous representation
$$
U:\1x\in \7R \mapsto U(\1x)\in\2U(\2H_0)
$$
of the 
spatial translations in $\Sgm\cong\7R$, such that 
$\al_{\1x}:=\Ad(U(\1x))$ maps $\6M(\2I)$ onto $\6M(\2I+\1x)$.
\edes

A one-parameter group of automorphisms 
$\al=\{\al_t\in\aut(\6M);t\in\7R\}$ ($\aut(\6M)$ denotes the automorphisms 
of $C^*(\6M)$) is called a {\em dynamics} of the net $\6M$ if the following 
conditions are fulfilled:
\bdes
\itno a 
The automorphism group $\al$ has {propagation speed} $\ps(\al)\leq 1$, where 
$\ps(\al)$ is defined by:
\beqa
\ps(\al):=
\inf\{\beta'|\al_t\6M(\2I)\subset \6M(\2I_{\beta'|t|});
\forall t,\2I\} \ \ .
\eeqa
Here $\2I_s:=\2I+(-s,s)$ denotes the interval, enlarged by $s>0$. 
\itno b 
The automorphisms $\{\al_t\in\aut(\6M);t\in\7R\}$ 
commute with the automorphism group 
of spatial translations $\{\al_\1x\in\aut(\6M);\1x\in\7R\}$, i.e.:
\beqa
\al_t\circ\al_\1x=\al_\1x\circ\al_t \ \ \ ; \ \ \ \forall \1x,t \  \ .
\eeqa
\edes
The set of all dynamics of $\6M$ is denoted by $\dyn(\6M)$.

For our purposes it is crucial to distinguish carefully the
C*-inductive limit $C^*(\6M)$ of the net $\6M$ and   
the corresponding C*- and W*-algebras, which belong to an 
unbounded region $\2J\subset\Sgm$. They are denoted by
$$
C^*(\6M,\2J):=\ol{\bigcup_{\2I\subset\2J}\6M(\2I)}^{||\cdot||}
\ \ \mbox{ and } \ \ 
\6M(\2J):=\bigvee_{\2I\subset\2J}\6M(\2I) \ \ \ \mbox{ respectively.}
$$
 
The Cauchy data of the $P(\phi)_2$- and the $\mbox{Yukawa}_2$
model are given by the nets of the corresponding free fields 
at time $t=0$. For these Cauchy data it can be proven that 
the inclusion 
\beqa
\6M(\2I_{RR})\subset \6M(\2I_R)
\eeqa 
is split \cite{Bu1,Sum1,Schl4}. 

Let us briefly explain how kink states can be constructed if  
the following conditions are assumed:
\bdes
\itno i
The dynamics of the model satisfies an appropriate 
{\em extendibility condition}
which we shall explain later.
\itno {ii}
The vacuum states are local Fock states which is 
automatically satisfied for $P(\phi)_2$ and $\mbox{Yukawa}_2$ models 
\cite{GlJa1,Schra1}. 
\edes
   
\paragraph*{\it Step 1':}
We consider the two-fold net 
\beqa
\6M\olt\6M:\2I\mapsto \6M(\2I)\olt\6M(\2I)  \ \ .
\eeqa
Like in {\it Step 1} 
of our previous construction scheme, the split property 
implies that on $\6M(\2I_{RR})\olt\6M(\2I_{RR})$, the flip automorphism 
is implemented by a unitary operator $\te_{\2I}$
\cite{AntLo}. The adjoint action of $\te_{\2I}$ 
is an automorphism $\beta^{\2I}$  
which has the following properties:  
\bdes
\itno i 
The automorphism $\beta^{\2I}$ acts trivially on observables which are 
localized in the left complement of $\2I$ and it acts like the 
flip on observables which are localized in the right complement of $\2I$. 
\itno {ii}
The automorphism $\beta^{\2I}$ maps local algebras into local algebras.
\edes
Note that the automorphism $\be^\2I$ does not depend on the dynamics $\al$. 

\paragraph*{\it Step 2':}
Let $\om_1$, $\om_2$ be two vacuum states with respect 
to a given dynamics $\al$. 
The state 
\beqa
\om:=\om_1\otimes\om_2\circ\beta^\2I|_{C^*(\6M)\otimes\7C\11}
\eeqa
interpolates the vacua $\om_1$ and $\om_2$. Moreover, it is 
covariant under spatial translations since for each $\1x$ the operator   
\beqa
\gam(0,\1x)=(\al_\1x\otimes\al_\1x)(\te_\2I)\te_\2I
\eeqa
is localized in a sufficiently large bounded interval. Indeed, 
the unitary operators  
\beqa
U(0,\1x):=(U_1(0,\1x)\otimes U_2(0,\1x)) \ (\pi_1\otimes\pi_2)(\gam(0,-\1x)) 
\eeqa
implement the spatial translations in the GNS-representation 
of $\om$ where $U_1$ and $U_2$ implement the translations in the
GNS-representations $\pi_1,\pi_2$ of $\om_1$ and $\om_2$ respectively. 

\paragraph*{\it Step 3':}
It remains to be proven that $\om$ is translationally 
covariant with respect to the dynamics $\al$. 
For this purpose, we wish to construct a cocycle  
$\gam(0,t)$ such that the operators 
\beqa
U(t,0):=(U_1(t,0)\otimes U_2(t,0)) \ (\pi_1\otimes\pi_2)(\gam(-t,0)) 
\eeqa
implement the dynamics $\al$ in the GNS-representation 
of $\om$. The operator 
\beqa
\gam(t,0):=(\al_t\otimes\al_t)(\te_\2I)\te_\2I
\eeqa
is a formal solution. Unfortunately, the flip implementer 
$\te_\2I$ is {\em not} contained in 
any local algebra and the term 
$(\al_t\otimes\al_t)(\te_\2I)$ has no mathematical meaning
unless $\al$ is the free dynamics. However, it can be given a meaning 
in some cases. We shall see that for an interacting dynamics 
there exists a suitable  
cocycle of the operators $\gam(t,0)$ such that $\gam(t,0)$
is localized in a bounded interval whose size depends 
linearly on $|t|$. 

In order to formulate a sufficient condition for the existence 
of $\gam(t,0)$, we construct an extension of the net 
$\6M\olt\6M$. 
We define $\hat\6M(\2I)$ to be the von Neumann algebra which is generated by 
$\6M(\2I)\olt\6M(\2I)$ and the operator $\te_\2I$. 
The net 
\beqa
\hat\6M:\2I\mapsto\hat\6M(\2I)
\eeqa
is an extension of $\6M\olt\6M$ which does not fulfill locality. This is 
due to the non-trivial implementation properties of $\te_\2I$.  
We shall call a dynamics $\al$ {\em extendible} if there exists a dynamics 
$\hat\al$ of $\hat\6M$ which is an extension of $\al\otimes\al$. Indeed,  
\beqa
t\mapsto \gam(t,0):=\hat\al_t(\te_\2I)\te_\2I
\eeqa
is a cocycle which has the desired properties.
Finally, we conclude like in {Step 3} of our previous construction scheme
that the state 
\beqa
\om:=\om_1\otimes\om_2\circ\beta^\2I|_{C^*(\6M)\otimes\7C\11}
\eeqa
is a kink state where $\om_1,\om_2$ are vacuum states 
with respect to the dynamics $\al$. 
 
Since the extendibility condition is rather technical one might worry
that it is only fulfilled for few exceptional cases.
Fortunately, this is not true. 
There is a large class of quantum field theory models 
whose dynamics are extendible. 
We shall prove that the extendibility holds for the following models:
\bdes
\itno i
$P(\phi)_2$-models. 
\itno {ii} 
$\mbox{Yukawa}_2$ models.
\itno {iii}
Special types of Wess-Zumino models.
\edes
Note that a Dirac spinor field 
contributes to the field content of 
the $\mbox{Yukawa}_2$ and Wess-Zumino models, and the 
nets of Cauchy data fulfill {\em twisted duality} 
instead of Haag duality \cite{Sum1}. According to recent results which 
have been established by M. M\"uger \cite{Mue1}, our results remain true 
for these cases also.  

Wess-Zumino models have been studied in several 
papers. We refer to the work of  
A. Jaffe, A. Lesniewski, J. Weitsman and S. Janowsky 
\cite{JanWeit0,JaLesWeit1,JaLesWeit2,JanWeit1,JaLes}.
It has been proven in \cite{JanWeit1} that some 
Wess-Zumino models possess more than one vacuum sector. 
An application of our construction scheme proves  
the existence of kink states for these models.

\section{Preliminaries}
\label{s3}

In the first part (Section \ref{s3a}) of this preliminary section, 
we briefly describe how to construct a Haag-Kastler net from a 
given net of Cauchy data and a given dynamics. 
Examples for physical states with respect to an 
interacting dynamics are given in the second part (Section \ref{s4}). 

\subsection{From Cauchy Data to Haag-Kastler Nets}\label{s3a}

We denote by $U(\6M)$ the group of unitary operators in $C^*(\6M)$.
Let $\6G(\7R,\6M)$ be the group which is generated by 
the set 
$$
\{ (t,u)|\mbox{ $t\in\7R$ and $u\in U(\6M)$ } \}
$$
modulo the following relations:
\bdes
\itno 1
For each $u_1,u_2\in U(\6M)$ and for each $t_1,t_2,t\in\7R$, we require: 
$$
(t,u_1)(t,u_2)=(t,u_1u_2) \mbox{ and } (t,\11)=\11
$$
\itno 2
For $u_1\in\6M(\2I_1)$ and $u_2\in\6M(\2I_2)$ with 
$\2I_1\subset(\2I_2+[-|t|,|t|])^c$ we require for each $t_1\in\7R$:
$$
(t_1+t,u_1)(t_1,u_2)=(t_1,u_2)(t_1+t,u_1) 
$$
\edes
We conclude from relation {\it (1)} that $(t,u)$ is the inverse 
of $(t,u^*)$. Furthermore, a localization 
region in $\7R\times\Sgm$ can be assigned
to each element in $\6G(\7R,\6M)$. 
A generator $(t,u)$, $u\in\6M(\2I)$ 
is localized in $\2O\subset\7R\times\Sgm$ if $\{t\}\times\2I\subset\2O$.
The subgroup of $\6G(\7R,\6M)$ which is generated by elements which are  
localized in the double cone $\2O$, is denoted by $\6G(\2O)$.
 
We easily observe that relation {\it (2)} implies that group elements commute 
if they are localized in space-like separated regions.  

The translation group in $\7R^2$ is naturally represented by  
group-automorphisms of $\6G(\7R,\6M)$. They are defined 
by the prescription  
$$
\beta_{(t,\8x)}(t_1,u):=(t+t_1,\al_\8xu) \ \ \ .
$$
Thus the subgroup $\6G(\2O)$ is mapped onto $\6G(\2O+(t,\8x))$
by $\beta_{(t,\8x)}$.

To construct the universal Haag-Kastler net, we build the group C*-algebra
$\6B(\2O)$ with respect to $\6G(\2O)$. For convenience, we shall 
describe the construction of $\6B(\2O)$ briefly.

In the first step we build the *-algebra $\6B_0(\2O)$ which is generated 
by all complex valued functions $a$ on $\6G(\2O)$, such that 
\beqa
a(u)=0 \ \ \mbox{ for almost each $u\in\6G(\2O)$ .} 
\eeqa
We write such a function symbolically as a formal sum, i.e.
\beqa
a=\sum_u a(u)\ u
\eeqa
The product and the *-relation is given as follows:
\beqa
ab= \sum_u a(u) \ u \cdot \sum_{u'} b(u') \ u' =
\sum_{u'}\bl(\sum_u a(u)b(u^{-1}u')\br) u'
\vs\vs
a^*=\sum_u \bar a(u^{-1}) \ u
\eeqa

It is well known, that the algebra $\6B_0(\2O)$ has a 
C*-norm which is given by
\beqa
||a||:=\sup_\pi||\pi(a)||_\pi
\eeqa
where the supremum is taken over each Hilbert space representation $\pi$ 
of $\6B_0(\2O)$. Finally, we define $\6B(\2O)$ as the closure of 
$\6B_0(\2O)$ with respect to the norm above.

The C*-algebra which is generated by all local algebras $\6B(\2O)$ 
is denoted by $C^*(\6B)$. 
By construction, the group isomorphisms $\beta_{(t,\8x)}$ 
induce a representation of the translation group by automorphisms 
of $C^*(\6B)$. 

\paragraph*{\it Observation:}
The net of C*-algebras
$$
\6B:=\{\6B(\2O)|\2O \ \mbox{ is a bounded double cone in $\7R^2$ }\}
$$ 
is a translationally covariant Haag-Kastler net. 
\skb

We have to mention that the universal 
net $\6B$ is not Lorentz covariant. The universal properties of the net $\6B$
are stated in the following Proposition:

\bpro\label{pro0}
Each dynamics $\al\in\dyn(\6M)$ induces a C*-homomorphism 
$$
\iota_\al:C^*(\6B)\to C^*(\6M)
$$
such that
$$
\iota_\al\circ\beta_{(t,\8x)}=\al_{(t,\8x)}\circ\iota_\al \ \ \mbox{ ,}
$$
for each $(t,\8x)\in\7R^2$.
In particular,   
$$
\6A_\al:\2O\mapsto\6A_\al(\2O):=\iota_\al(\6B(\2O))'' 
$$
is a translationally covariant Haag-Kastler net.
\epro
\bpr
Given a dynamics $\al$ of $\6M$.
We conclude from $\ps(\al)\leq 1$ that the prescription 
$$
(t,u)\mapsto \al_tu  
$$
defines a C*-homomorphism  
$$
\iota_\al:C^*(\6B)\to C^*(\6M)\ \ .
$$
In particular, $\iota_\al$ is a representation of $C^*(\6B)$
on the Hilbert space $\2H_0$. 
This statement can be obtained by using the relations, listed below.
\bdes
\itno a
$$
\iota_\al((t,u_1)(t,u_2))=\al_tu_1\al_tu_2=\al_t(u_1u_2)=
\iota_\al(t,u_1u_2)
$$
\itno b
If $(t_1,u_1)$ and $(t_1+t,u_2)$ are localized in space-like separated 
regions, then we obtain from $\ps(\al)\leq 1$:
$$
[\iota_\al(t_1,u_1),\iota_\al(t_1+t,u_2)]=
\al_{t_1}[u_1,\al_tu_2]=0
$$
\itno c
$$
\iota_\al(\beta_{(t,\8x)}(t_1,u))=\iota_\al(t+t_1,\al_{\8x}u)=
\al_{(t,\8x)}\al_{t_1}u
$$
\edes
\epr

In general we expect that for a given dynamics $\al$ the representation 
$\iota_\al$ is not faithful. Hence each dynamics defines a two-sided 
ideal 
$$
J(\al):=\iota_\al^{-1}(0)\in C^*(\6B)
$$
in $C^*(\6B)$ which we call the {\em dynamical ideal} 
with respect to $\al$ and the quotient C*-algebras 
\beqa
\6B(\2O)/J(\al)\cong\6A_\al(\2O)
\eeqa
may depend on the dynamics $\al$. Indeed, if $\2O$ is a double cone 
whose base is {\em not contained} in $\Sgm$, then 
for different dynamics $\al_1,\al_2$ the algebras $\6A_{\al_1}(\2O)$ and
$\6A_{\al_2}(\2O)$ are different. On the other hand, if the base 
of $\2O$ is contained in $\Sgm$, then we conclude from 
the fact that the dynamics $\al$ has finite propagation speed and 
from Proposition \ref{pro0}:
\bcor
If $\2I\subset\Sgm$ is the base of the double cone $\2O$, then the algebra
$\6A_\al(\2O)$ is independent of $\al$.
In particular, the C*-algebra 
$$
C^*(\6M)
=\ol{\bigcup_\2I \6M(\2I)}^{||\cdot||} 
=\ol{\bigcup_\2O \6A_\al(\2O)}^{||\cdot||} 
$$
is the C*-inductive limit of the net $\6A_\al$.
\ecor
From the discussion above, we see that two dynamics with the same dynamical 
ideal induces the same quantum field theory. 

\subsection{Examples for Physical States}
\label{s4}

Let us consider the set $\6S$ of all {\em locally normal states}
on $C^*(\6M)$, i.e. for each state $\om\in\6S$ and for each bounded interval
$\2I$, the restriction
$$
\om|_{\6M(\2I)}
$$
is a normal state on $\6M(\2I)$. 

As mentioned in the introduction, we are interested in 
states with vacuum and particle-like properties, i.e. 
states satisfying the {\em Borchers criterion} 
(See the Introduction for this notion).

\paragraph*{\it Notation:}
Given a dynamics $\al\in\dyn(\6M)$. We denote the corresponding 
set of all locally normal
states which satisfies the Borchers criterion 
by $\6S(\al)$ and analogously the set of all vacuum states by $\6S_0(\al)$.
Moreover, we write for the set 
of vacuum sectors 
\beq
\sec_0(\al):=\{[\om]|\om\in\6S_0(\al) \}
\eeq
where $[\om]$ denotes the unitary equivalence class of the 
the GNS-representation of $\om$. 
\paragraph*{}

\paragraph{Examples:}
Examples for vacuum states are the vacua of 
the $P(\phi)_2$-models \cite{GlJa1,GlJa2}. 
The interacting part of the cutoff Hamiltonian is 
given by a Wick polynomial of the time zero field $\phi_0$, i.e.
$$
H_1(\2I)=H_1(\chi_\2I)=:P(\phi_0):(\chi_\2I)
$$
where $\chi_\2I$ is a test function with $\chi_\2I(\1x)=1$ for 
$\1x\in\2I$ and $\chi_\2I(\1y)=0$ 
on the complement of a slightly lager region $\hat\2I\supset\2I$.
It is well known that $H_1(\2I)$ is a self-adjoint operator, which 
has a joint core with the free Hamiltonian $h_0$, and is affiliated with
$\6M(\hat\2I)$.
The operator $h_1(\2I)$ induces a automorphism group 
$\al_\2I$ which is given by
$$
\al_{\2I,t}(a):=e^{iH_1(\2I)t}ae^{-iH_1(\2I)t} \ \ \ .
$$
Consider the inclusion of intervals $\2I_0\subset\2I_1\subset\2I_2$.
Then we have for each $a\in\6M(\2I_0)$:
$$
\al_{\2I_1,t}(a)=\al_{\2I_2,t}(a)
$$
Hence, there exists a one-parameter automorphism group \newline
$\{\al_{1,t}\in\aut(\6M);t\in\7R\}$ such that 
$\al_{1,t}$ acts on $a\in\6M(\2I)$ as follows:
$$
\al_{1,t}(a)=\al_{\2I,t}(a) \ \ \ ; \ \ \ \forall t\in\7R
$$
The automorphism group $\{\al_{1,t}\in\aut(\6M);t\in\7R\}$ is a 
dynamics of $\6M$ with zero propagation speed, i.e. $\ps(\al_1)=0$.

Since $H_1(\2I)$ has a joint core with the free Hamiltonian $H_0$,
we are able to define the Trotter product of the automorphism groups
$\al_0$ and $\al_1$ which is given 
for each local operator $a\in\6M(\2I)$ by
\beqa
\al_t(a):=(\al_0\times\al_1)_t(a)=
s-\lim_{n\to\infty}(\al_{0,t/n}\circ\al_{1,t/n})^n(a)  \ \ \ .
\eeqa
The limit is taken in the strong operator topology.
Furthermore,the propagation speed is sub-additive with respect to the 
Trotter product \cite{GlJa1}, i.e.
$$
\ps(\al_0\times\al_1)\leq\ps(\al_0)+\ps(\al_1)
$$
and we conclude that 
$\al\in\dyn(\6M)$ is a dynamics of $\6M$. We call the dynamics $\al$ 
{\em interacting}. 

It is shown by Glimm and Jaffe \cite{GlJa1}
that there exist vacuum states 
$\om$ with respect to the interacting dynamics $\al$.
We have to mention, that there is {\em no} vector $\psi$ in 
Fock space $\2H_0$, such that the state
\beqa
a\mapsto\<\psi,a\psi\>
\eeqa
is a vacuum state with respect to an interacting dynamics $\al$, but there
is a sequence of vectors $(\Om_n)$ in $\2H_0$ such that the 
weak* limit
\beqa
\om=w^*-\lim_n \<\Om_n,\cdot\Om_n\>
\eeqa
is a vacuum state with respect to the dynamics $\al$.

\section{On the Existence of Kink States}
\label{s6}

The main theorem (Theorem \ref{the1}) 
of this paper is formulated in the first part 
(Section \ref{s6a}) of the present section. 
In order to prepare the proof of Theorem \ref{the1}, we 
need some technical preliminaries which are given in 
Section \ref{s6b}. In the last part (Section \ref{s6c}), 
we prove a criterion for the existence of kink states 
(the extendibility of the dynamics) which turns out 
to be satisfied by the $P(\phi)_2$ and Yukawa$_2$ models.

\subsection{The Main Result}
\label{s6a}
We now reformulate the definition (see Introduction) 
of a kink state within the time-slice formulation. 

\bdef\label{def4}
Let $\al\in\dyn(\6M)$ be a dynamics of $\6M$.  
A state $\om$ of $\6M$ is called a {\em kink state}, 
interpolating vacuum states
$\om_1,\om_2\in\6S_0(\al)$ if 
\begin{description}
\item[{\it (a)}]
$\om$ satisfies the Borchers criterion
\item[{\it (b)}]
and there exists a bounded interval $\2I$, such that $\om$ 
fulfills the relations:
$$
\pi|_{C^*(\6M,\2I_{LL})}\cong\pi_1|_{C^*(\6M,\2I_{LL})} \ \ \mbox{ and } \ \ 
\pi|_{C^*(\6M,\2I_{RR})}\cong\pi_2|_{C^*(\6M,\2I_{RR})}    
$$
where the symbol $\cong$ means unitarily equivalent
and $(\2H,\pi,\Om),(\2H_j,\pi_j,\Om_j)$ are the GNS-triples of
the states $\om\in\6S(\al)$ and $\om_j\in\6S_0(\al);j=1,2$ respectively.
\end{description}
The set of all kink states which interpolate $\om_1$ and $\om_2$ is denoted by
\newline
$\6S(\al|\om_1,\om_2)$.
\eef

As already mentioned in the Introduction, a criterion 
for the existence of an interpolating kink state,
can be obtained by looking at the construction method of \cite{Schl4}.
In our context, we have to select a class of dynamics which are 
equipped with {\em good properties}. Such a selection criterion is 
developed in section \ref{s6}.
We shall show that each 
dynamics of a $P(\phi)_2$-model satisfies this criterion which leads 
to the following result:

\bthe\label{the1}
If $\al\in\dyn(\6M)$ is a dynamics of a model with 
$P(\phi)_2$ plus Yukawa$_2$ interaction, then for each 
pair of vacuum states $\om_1,\om_2\in\6S_0(\al)$ there exists an 
interpolating kink state $\om\in\6S(\al|\om_1,\om_2)$.
\ethe

We shall prepare the  proof of Theorem \ref{the1} during the subsequent 
sections.

\subsection{Technical Preliminaries}\label{s6b}

\bdef
Let $\6M$ be a net of Cauchy data.
We denote by $G(\6M)$ the group of unitary operators $u\in\6B(\2H_0)$ 
whose adjoint actions $\chi_u:=\Ad(u)$ commute with 
the spatial translations, i.e.:
\beqa
\chi_u\circ\al_\1x=\al_\1x\circ\chi_u \ \ .
\eeqa
Let $\al\in\dyn(\6M)$ be a dynamics of the net $\6M$. Then we define 
the following sub-group of $G(\6M)$:
\beqa
G(\al,\6M):=\{u\in G(\6M)|\chi_u\circ\al_t=\al_t\circ\chi_u \ 
\mbox{ for each $t\in\7R$.}\}
\eeqa
\eef

\brem
Each operator $u\in G(\al,\6M)$ induces a 
symmetry of the Haag-Kastler net $\6A_\al$.
\erem

We make the following assumptions for the 
net of Cauchy data $\6M$:

\paragraph*{\it Assumption:}
\bdes
\itno a
The net $\6M$ fulfills duality, i.e.
\beq
\6M(\2I)'=\6M(\2I_{LL})\vee\6M(\2I_{RR})
\eeq
\itno b
There exists a dynamics $\al_0$ and a normalized vector $\Om_0$ in $\2H_0$, 
such that 
\beqa
\om_0=\<\Om_0,(\cdot)\Om_0\> 
\eeqa
is a vacuum state with 
respect to the dynamics $\al_0$. 
\itno c
For each bounded interval $\2I$, the inclusion
\beqa
(\6M(\2I_{RR}),\6M(\2I_R))
\eeqa
is split.
\itno d
The net fulfills weak additivity.
\edes
\skb

According to our assumption, we conclude from 
the Theorem of Reeh and Schlieder that 
$\Om_0$ is a standard vector for the inclusion 
$(\6M(\2I_{RR}),\6M(\2I_R))$ which implies that 
\beq\label{splitincl}
\Lam(\2I):=(\6M(\2I_{RR}),\6M(\2I_R),\Om_0)
\eeq
is a standard split inclusion for each interval $\2I$.
and hence (see \cite{DLo}) 
there exists a unitary operator 
$$
w_{\2I}:\2H_0\otimes \2H_0\to  \2H_0
$$
such that for $a\in \6M(\2I_{LL})$ and $b\in\6M(\2I_{RR})$ we have: 
$$
w_{\2I}(a\otimes b)w_{\2I}^*=ab
$$
Thus there is an interpolating type I factor $\2N\cong\6B(\2H_0)$, i.e.
$$
\6M(\2I_{RR})\subset\2N\subset \6M(\2I_R)
$$
which is given by 
$$
\2N:=w_{\2I}(\11\otimes \6B(\2H_0))w_{\2I}^* \ \ \ .
$$
Hence we obtain an embedding of $\6B(\2H_0)$ into the algebra 
$\6M(\2I_R)$:
$$
\Psi_\2I:F\in\6B(\2H_0)\mapsto 
w_{\2I}(\11\otimes F)w_{\2I}^*\in\6M(\1x,\infty)
$$
This embedding is called the {\em universal localizing map}.

\brem
We shall make a few remarks on the assumptions given above.
\bdes
\itno {i}
The results, which we shall establish in the following, remain
correct if the 
net of Cauchy data fulfills 
{\em twisted duality} instead of duality \cite{Mue1,Sum1}. 
\itno {ii}
For the application of our analysis to quantum field theory models, 
like $P(\phi)_2$- or $\mbox{Yukawa}_2$ models, 
we can choose as Cauchy data tensor products of the time-zero algebras of the 
massive free Bose or Fermi field. 
The time-zero algebras of the massive free Bose field 
fulfill the assumptions {\it (a)} and {\it (b)}
and it has been shown  \cite[Appendix]{Schl5} (compare also \cite{Bu1}) that 
{\it (c)} is also fulfilled. 
Replacing duality by twisted duality,   
the assumptions {\it (a)} to {\it (c)} hold for the
massive free Fermi field, too \cite{Sum1}.
In addition to that, we claim that the weak additivity {\it (d)} 
is also fulfilled in these cases.
\itno {iii}
The state $\om_0$ plays the role of a free massive vacuum state, 
called the bare vacuum.  
\edes
\erem

\bpro\label{pro51}
Let $u\in G(\6M)$ be an operator and let $\2I$ be a bounded interval.
Then there exists a canonical automorphism $\chi_u^\2I$ with the properties:
\bdes
\itno 1
The relations
\beq\label{inpl2}
\chi_u^\2I|_{C^*(\6M,\2I_{LL})}=\id_{C^*(\6M,\2I_{LL})} \ \ \mbox{ and} \ \
\chi_u^\2I|_{C^*(\6M,\2I_{RR})}=\chi_u|_{C^*(\6M,\2I_{RR})} 
\eeq
hold.
\itno 2
There exists a strongly continuous map 
$\gam^1_{(u,\2I)}:\Sgm\to C^*(\6M)$ such that: 
\bdes
\itno i 
\beqa
\Ad(\gam^1_{(u,\2I)}(\1x))=
\al_\1x\circ\chi_u^\2I\circ\al_{-\1x}\circ(\chi_u^\2I)^{-1} \ \ .
\eeqa
\itno {ii}
The cocycle condition is fulfilled:
\beqa
\gam^1_{(u,\2I)}(\1x+\1y)=\al_\1x(\gam^1_{(u,\2I)}(\1y))
\gam^1_{(u,\2I)}(\1x) \ \ .
\eeqa 
\edes
\edes
\epro
\bpr\bdes
\itno 1
In the same manner as in \cite{Schl4}, we show that 
\beqa
\Ad(\Psi_\2I(\11\otimes u))(\6M(\hat\2I))\subset\6M(\hat\2I)
\eeqa
if the interval $\hat\2I$ contains $\2I$.
This implies that  
\beqa
\chi_u^\2I:=\Ad(\Psi_\2I(\11\otimes u))
\eeqa
is a well defined automorphism of $C^*(\6M)$.
By using the properties of the universal localizing map $\Psi_\2I$, 
we conclude that $\chi_u^\2I$ fulfills equation (\ref{inpl2}).
\itno 2
By a straight forward generalization of 
the proof of \cite[Proposition 4.2]{Schl4}, 
we conclude that the statement {\it (2)} holds where 
$\gam^1_{(u,\2I)}(\1x)$ is given by:
\beqa
\gam^1_{(u,\2I)}(\1x)=\Psi_{\2I+\1x}(\11\otimes u)\Psi_\2I(\11\otimes u^*)\ \ .
\eeqa 
\edes
\epr

Let $\om$ be a vacuum state with respect to the dynamics $\al$ and let 
$u\in G(\al,\6M)$, then 
the state 
\beqa
\om_u^\2I:=\om\circ\chi_u^\2I
\eeqa
seem to be a good candidate for an interpolating kink state.
Indeed, it follows from the construction of $\chi_u^\2I$ that 
\beqa
\om_u^\2I|_{C^*(\6M,\2I_{RR})}&=&\om\circ\chi_u|_{C^*(\6M,\2I_{RR})}
\vs\vs
\om_u^\2I|_{C^*(\6M,\2I_{LL})}&=&\om|_{C^*(\6M,\2I_{LL})} \ \ .
\eeqa
Hence $\om_u^\2I$ interpolates $\om$ and $\om\circ \chi_u$. 
To decide whether 
$\om_u^\2I$ is a positive energy state, we  
investigate in the subsequent section, 
how $\chi_u^\2I$ is transformed under the action of a dynamics $\al$.

\subsection{When Does a Theory Possess Kink States?}
\label{s6c}

Let $\al$ be a dynamics and 
$G\subset G(\al,\6M)$ be a {\em finite} subgroup.  
By using the universal localizing map $\Psi_\2I$, 
we obtain for each bounded interval $\2I$ a unitary representation 
of $G$ 
\beqa
U_\2I:G\ni g \mapsto U_\2I(g):=\Psi_\2I(\11\otimes g)\in \6M(\2I_R) \ \ .
\eeqa

In the previous section it has been shown that 
$U_\2I(g)$ implements an automorphism $\chi^\2I_g$ which is 
covariant under spatial translations (Proposition \ref{pro51}).
For a dynamics $\al\in\dyn(\6M)$, we wish to construct a cocycle  
$\gam_{(g,\2I)}$ in order to 
show that $\chi^\2I_g$ is an interpolating automorphism. 
The {\em formal} operator 
\beqa
\gam_{(g,\2I)}(t,\1x):=\al_{(t,\1x)}(U_\2I(g))U_\2I(g)^*
\eeqa
seems to be a useful Ansatz since it formally implements the automorphism 
\beqa
\al_{(t,\1x)}\circ\chi^\2I_g\circ\al_{(-t,-\1x)}\circ (\chi^\2I_g)^{-1} \ \ .
\eeqa
Unfortunately, the operators $U_\2I(g)$ are {\em not} contained in $C^*(\6M)$
and the term \newline $\al_{(t,\1x)}(U_\2I(g))$ 
has no well defined mathematical meaning.
To get a well defined solution for $\gam_{(g,\2I)}$, 
we construct an extension of the net 
$\6M$ which contains the operators $U_\2I(g)$ (compare also \cite{Mue1}).

\bdef\label{exten}
Let $G\subset G(\6M)$ be a compact sub-group. The net 
$\6M\rtimes G$ is defined by the assignment
\beqa
\6M\rtimes G:\2I\mapsto (\6M\rtimes G)(\2I):=\6M(\2I)\vee U_{\2I}(G)'' \ \ .
\eeqa
\eef

\bpro\label{pro52}
Let $\2I$ be a bounded interval, then the map
\beqa
\pi^\2I:\6M(\2I)\rtimes G\ni a \cdot g \mapsto 
a \ U_{\2I}(g)\in\6M(\2I)\vee U_{\2I}(G)''
\eeqa
is a faithful representation of the crossed product $\6M(\2I)\rtimes G$.
\epro
\bpr
First, we easily observe that $\pi^\2I$ is a well defined representation
of \newline $\6M(\2I)\rtimes G$. According to 
\cite[Theorem 2.2, Corollary 2.3]{Hag}, we conclude that the 
crossed product $\6M(\2I)\rtimes G$ is isomorphic 
to the von Neumann algebra $\6M(\2I)\vee U_{\2I}(G)''$ and 
$\pi^\2I$ is a W*-isomorphism. 
\epr

\bdef
A one parameter
automorphism group $\al$, which satisfies the conditions, listed below, 
is called a $G$-dynamics of the extended net $\6M\rtimes G$.
\bdes
\itno a
$\al$ is a dynamics of the net $\6M\rtimes G$ (See Introduction).
\itno b
The automorphisms $\al_t$ commute with the automorphisms $\chi_g$, i.e.
\beqa
\al_t\circ\chi_g=\chi_g\circ\al_t \ \ ; \mbox{ for each $t\in\7R$ and 
for each $g\in G$.}
\eeqa
\edes
The set of all 
$G$-dynamics of $\6M\rtimes G$ is denoted by $\dyn_G(\6M\rtimes G)$.
\eef

\bpro\label{pro521} 
Let $\al\in\dyn_G(\6M\rtimes G)$ be a $G$-dynamics and
$\2I$ be a bounded interval.
Then the operator  
\beqa
\gam^0_{(g,\2I)}(t):=\al_t(U_\2I(g))U_\2I(g)^*
\eeqa
is contained in $\6M(\2I_{|t|})$ where $\2I_{|t|}$ denotes the 
enlarged interval $\2I+(-|t|,|t|)$ and the operator 
\beqa
\gam_{(g,\2I)}(t,\1x):=\al_{(t,\1x)}(U_\2I(g))U_\2I(g)^*
\eeqa
fulfills the cocycle condition:
\beqa
\gam_{(g,\2I)}(t+t',\1x+\1x')=
\al_{(t,\1x)}(\gam_{(g,\2I)}(t',\1x'))\gam_{(g,\2I)}(t,\1x) \ \ .
\eeqa
\epro
\bpr
For $a\in C^*(\6M,\2I_{|t|,RR})$, the operator 
$\al_{-t}(a)$ is contained in $C^*(\6M,\2I_{RR})$ which implies
\beqa
a \ \al_t(U_\2I(g))U_\2I(g)^* &=& \al_t(\al_{-t}(a)U_\2I(g))U_\2I(g)^*
\vs\vs
&=& \al_t(U_\2I(g)\chi_g\al_{-t}(a))U_\2I(g)^* 
\vs\vs
&=& \al_t(U_\2I(g)\al_{-t}\chi_g(a))U_\2I(g)^* 
\vs\vs
&=& \al_t(U_\2I(g))\chi_g(a)U_\2I(g)^*
\vs\vs
&=& \al_t(U_\2I(g))U_\2I(g)^* \ a
\eeqa
and we conclude:
\beqa
\al_t(U_\2I(g))U_\2I(g)^*\in C^*(\6M,\2I_{|t|,RR})'=\6M(\2I_{|t|,L})
\eeqa
By a similar argument, $\al_t(U_\2I(g))U_\2I(g)^*$ is contained
in $\6M(\2I_{|t|,R})$ and we conclude from duality that 
it is contained in $\6M(\2I_{|t|})$. 
The cocycle condition for $\gam_{(g,\2I)}$ is obviously fulfilled
and the proposition follows.
\epr

\bdef
Let $\al\in\dyn(\6M)$ be a dynamics and $G\subset G(\6M)$ be 
a compact subgroup.
We shall call $\al$ {\em $G$-extendible}
if there exists a $G$-dynamics $\hat\al$ of the extended net 
$\6M\rtimes G$, such that 
\beqa
\hat\al_t|_{C^*(\6M)}=\al_t
\eeqa
for each $t\in\7R$.
\eef

We are now prepared to prove one of our key results:
\bthe\label{the51}
Let $\al\in\dyn(\6M)$ be a $G$-extendible dynamics and let $\chi^\2I_g$ be
the automorphism which can be constructed by 
Proposition \ref{pro51}. Then for each vacuum state 
$\om$ with respect to $\al$ the state 
\beqa
\om_g^\2I:=\om\circ\chi_g^\2I
\eeqa
is a kink state which interpolates $\om$ and $\om\circ\chi_g$.
\ethe
\bpr
As postulated, there exists an extension 
$\hat\al\in\dyn_G(\6M\rtimes G)$ of $\al$.
We show that for each $g\in G$ the operator 
\beqa
\gam^0_{(g,\2I)}(t):=\hat\al_t(U_\2I(g))U_\2I(g)^*
\eeqa
implements the automorphism 
\beqa
\al_t\circ\chi^\2I_g\circ\al_{-t}\circ (\chi^\2I_g)^{-1} 
\eeqa
on $C^*(\6M)$. Indeed, we have for each $a\in C^*(\6M)$:
\beqa
\Ad(\gam^0_{(g,\2I)}(t))a &=& \hat\al_t(U_\2I(g))U_\2I(g)^* \ a \
U_\2I(g)\hat\al_t(U_\2I(g))^*
\vs\vs
&=&
\hat\al_t(U_\2I(g)) \ (\chi^\2I_g)^{-1}(a) \ \hat\al_t(U_\2I(g))^*
\vs\vs
&=&
\hat\al_t\biggr ( U_\2I(g)\al_{-t}
         \biggr ( (\chi^\2I_g)^{-1}(a) \biggl )U_\2I(g)^*\biggl )
\vs\vs
&=&
\al_t\biggr ( U_\2I(g)\al_{-t}
     \biggr ( (\chi^\2I_g)^{-1}(a) \biggl )U_\2I(g)^*\biggl )
\vs\vs
&=&
\al_t\circ\chi^\2I_g\circ\al_{-t}\circ(\chi^\2I_g)^{-1}(a)
\eeqa
Finally we conclude from Proposition \ref{pro521} that  
\beqa
\gam_{(g,\2I)}(t,\1x):=\al_{(t,\1x)}(U_\2I(g))U_\2I(g)^*
\eeqa
is a cocycle where $\gam_{(g,\2I)}(t,\1x)$ is localized in a 
sufficiently large bounded interval.
By a straight forward generalization of the proof 
of \cite[Proposition 5.5]{Schl4} we conclude that 
$\om_u^\2I$ is a positive energy state which implies the result. 
\epr

The dynamics $\al$ of $P(\phi)_2$- and $\mbox{Yukawa}_2$ models are locally 
implementable by unitary operators. More precisely, 
for each bounded interval $\2I$ and for each positive 
number $\tau>0$, there exists a unitary 
operator $u(\2I,\tau|t)$ with the properties:
\bdes
\itno 1
If $|t_1|,|t_2|,|t_1+t_2|<\tau$, then we have 
\beqa
u(\2I,\tau|t_1+t_2)=u(\2I,\tau|t_1)u(\2I,\tau|t_2) \ \ .
\eeqa
\itno 2
For $|t|<\tau$, the operator 
$u(\2I,\tau|t)$ implements $\al_t$ on $\6M(\2I)$, i.e.:
\beq\label{locimpl}
\al_t(a)=u(\2I,\tau|t) \ a \ u(\2I,\tau|t)^* \ \ ; \ 
\mbox{  for each $a\in\6M(\2I)$.} 
\eeq
\edes
Let $G\subset G(\al,\6M)$ be a compact sub-group. In order to show 
that $\al$ is $G$-extendible, it is sufficient to prove 
that the operators 
\beqa
u(\2I_1,\tau|t)U_\2I(g)u(\2I_1,\tau|t)^* \ \ ,
\eeqa
which are the obvious candidates 
for $\hat\al(U_I(g))$, 
are independent of $\2I_1$ for $\2I_1\supset\2I$ and $|t|\leq\tau$.

\blem\label{lem522}
If for each $\2I\subset\2I_1$, for each $\tau<\tau_1$ and for each $g\in G$
the equation   
\beq\label{extendyn}
u(\2I,\tau|t)U_\2I(g)u(\2I,\tau|t)^*
=u(\2I_1,\tau_1|t)U_\2I(g)u(\2I_1,\tau_1|t)^*
\eeq
holds, then the dynamics $\al$ is $G$-extendible. 
Here $u(\2I,\tau|t)$ are unitary operators which 
fulfill equation (\ref{locimpl}).
\elem
\bpr
Let $(\2I_n,\tau_n)_{n\in\7N}$ be a sequence, such that 
$\lim_n\2I_n=\7R$ and $\lim_n\tau_n=\infty$.  
We conclude from our assumption (equation (\ref{extendyn})) 
that the uniform limit 
\beqa
\hat\al_t(a):=\lim_{n\to\infty}\Ad(u(\2I_n,\tau_n|t))(a)
\eeqa
exists. Thus $\hat\al:t\mapsto\hat\al_t$ is a 
well defined one-parameter automorphism group, extending the dynamics $\al$.
It remains to be proven that $\hat\al$ has propagation speed 
$\ps(\hat\al)\leq 1$. Since $\hat\al$ is an extension 
of $\al$ and $\ps(\al)\leq 1$, we conclude for each 
$a\in C^*(\6M,\2I_{|t|,RR})$ and for each $b\in C^*(\6M,\2I_{|t|,LL})$:
\beqa
ab \ \hat\al_t(U_\2I(g)) 
&=&\hat\al_t(\al_{-t}(a)\al_{-t}(b)U_\2I(g))
\vs\vs
&=&\hat\al_t(U_\2I(g)\al_{-t}\chi_g(a)\al_{-t}(b)) 
\vs\vs
&=&\hat\al_t(U_\2I(g)) \ \chi_g(a)b
\eeqa
Thus the operator $\hat\al_t(U_\2I(g))$ 
is contained in $\6M(\2I_{|t|,R})$
and implements $\chi_g$ on \newline $\6M(\2I_{|t|,RR})$.
This finally implies:
\beqa
\hat\al_t(U_\2I(g))U_{\2I_{|t|}}(g)^*\in\6M(\2I_{|t|})
\eeqa
and the lemma follows. 
\epr

Let us consider the two-fold W*-tensor product of the net 
of Cauchy data, i.e.: 
\beqa
\6M\olt\6M:\2I\mapsto\6M(\2I)\olt\6M(\2I)
\eeqa

\bobs
\bdes
\itno {i}
If the net $\6M$ fulfills the conditions 
{\it (a)} to  {\it (c)} of the previous section, then the  
net $\6M\olt\6M$ fulfills them, too.
\itno {ii}
Let $\al\in\dyn(\6M)$ be a dynamics of $\6M$, then $\al^{\otimes 2}$ 
is a dynamics of $\6M\olt\6M$. Note that the flip operator 
$u_F$, which is given by 
\beqa
u_F:\2H_0\otimes\2H_0\to\2H_0\otimes\2H_0 \ \ ; \ \ 
\psi_1\otimes\psi_2\mapsto \psi_2\otimes\psi_1
\eeqa
is contained in $G(\al^{\otimes 2},\6M\olt\6M)$. Hence $u_F$ 
induces an embedding of $\7Z_2$ into $G(\al^{\otimes 2},\6M\olt\6M)$.
\itno {iii}
According to Definition \ref{exten}, we can construct 
a non-local extension 
\beqa
\hat\6M:=(\6M\olt\6M)\rtimes\7Z_2
\eeqa
of the two-fold net $\6M\olt\6M$. Let $\Psi_\2I$ be the 
universal localizing map of the 
standard split inclusion
\beqa
\Lam(\2I)\otimes\Lam(\2I)=
(\6M(\2I_{RR})^{\olt 2},\6M(\2I_R)^{\olt 2},\Om_0\otimes\Om_0)
\eeqa
and define $\te_\2I:=\Psi_\2I(\11\otimes u_F)$. 
Then the algebra $\hat\6M(\2I)$ is simply given by
\beqa
\hat\6M(\2I)=((\6M\olt\6M)\rtimes\7Z_2)(\2I)
=(\6M(\2I)\olt\6M(\2I))\vee \{\te_\2I\}'' \ \ \ .
\eeqa
\itno {iv}
By Proposition \ref{pro51}, there exists a canonical automorphism 
\beq\label{intauto}
\be^\2I:=\Ad(\te_\2I) \ \ ,
\eeq
associated with the pair $(u_F,\2I)$.
\edes
\eobs

\bnoa
Let $\al$ be a dynamics of $\6M$. In the sequel, we shall call $\al$ 
{\em extendible} if $\al^{\otimes 2}$ is $\7Z_2$-extendible.
\enoa

We conclude this section by 
the following corollary which can be derived by a direct application
of Theorem \ref{the51}:

\bcor\label{cor52}
Let $\al\in\dyn(\6M)$ be an extendible dynamics, then 
for each pair of vacuum states $\om_1,\om_2\in\6S_0(\6A_\al)$,
the state 
\beqa
\om=\mu_{\be^\2I}(\om_1\otimes\om_2) 
\eeqa
is a kink state.
\ecor

\section{Application to Quantum Field Theory Models}
\label{s52}

We show that a sufficient condition for the existence of interpolating 
automorphisms, i.e. the extendibility of the dynamics, is  
satisfied for the $P(\phi)_2$, the $\mbox{Yukawa}_2$ and special types
of Wess-Zumino models.

\subsection{Kink States in $P(\phi)_2$-Models}
\label{s521}

We shall show that the dynamics of $P(\phi)_2$-models are extendible.
As described in Section \ref{s4} the dynamics of a $P(\phi)_2$-model
consists of two parts.
\begin{description}
\item[{\it (1)}]
The first part is given by the free dynamics $\al_0$, with 
propagation speed \newline $\ps(\al_0)=1$, 
$$
\al_{0,t}(a)=e^{iH_0t}ae^{-iH_0t}
$$
where $(H_0,D(H_0))$ is the free Hamiltonian  
which is a self-adjoint operator
on the domain $D(H_0)\subset \2H_0$.
\item[{\it (2)}]
The second part is a dynamics $\al_1$ with propagation speed $\ps(\al_1)=0$, 
i.e. $\al_{1,t}$ maps each local algebra $\6M(\2I)$ onto itself.
The interaction part of the full Hamiltonian is given by 
a Wick polynomial of the time-zero field $\phi$:
$$
H_1(\2I)=H_1(\chi_\2I)=\int\8d\1x \ :P(\phi(\1x)): \ \chi_\2I(\1x)
$$
where $\chi_\2I$ is a smooth test function which is one on 
$\2I$ and zero on the complement of a 
slightly lager region $\hat\2I\supset\2I$.
The unitary operator $\exp(itH_1(\2I))$ implements the dynamics 
$\al_1$ locally, i.e. for each $a\in\6M(\2I)$ we have:
$$
\al_{1,t}(a):=e^{iH_1(\2I)t}ae^{-iH_1(\2I)t} \ \ .
$$
\end{description}

\bdef
An operator valued distribution $v:S(\7R)\to L(\2H_0)$ is called 
an {\em ultra local interaction}, if the following conditions are fulfilled:
\bdes
\itno 1
For each real valued test function $f\in S(\7R)$,
$v(f)$ is self-adjoint and has a common core with $H_0$.
\itno 2
Let $f\in S(\7R)$ be a real valued test function with support in a  
bounded interval $\2I$, then the spectral projections of $v(f)$
are contained in $\6M(\2I)$.
\itno 3
For each pair of test functions $f_1,f_2\in S(\7R)$, the 
spectral projections of $v(f_1)$ commute with the spectral 
projections of $v(f_2)$.
\edes
\eef

\brem
It has been proven in \cite{GlJa1}, that the Wick polynomials 
of the time zero fields are ultra local interactions.
Furthermore, each ultra local 
interaction $v$ induces a dynamics $\al^v\in\dyn(\6M)$
with propagation speed $\ps(\al^v)=0$. Let $\2I$ be a bounded 
interval and let $\chi_\2I\in S(\7R)$ be a positive test function with 
$\chi_\2I(\1x)=1$ for each $\1x\in\2I$. Indeed, 
by an application of J. Glimm's and A. Jaffe's analysis \cite{GlJa1}, 
we conclude that the automorphisms
\beqa
\al^v_t:\6M(\2I)\to\6M(\2I) \ \ 
; \ \ a\mapsto \Ad(\ \exp(itv(\chi_\2I)) \ ) a
\eeqa
define a dynamics with zero propagation speed.
In the sequel, 
we shall call a dynamics $\al^v$ {\em ultra local} if it is induced 
by an ultra local interaction $v$.
\erem

In order to prove that a dynamics $\al$, which is given by 
the Trotter product 
\beqa
\al=\al_0\times\al^v
\eeqa
of a free and an ultra local dynamics, is extendible, we show  
that each part of the dynamics can be extended separately. 

Since the free part of the dynamics can be extended to the algebra $\6B(\2H_0)$
of all bounded operators on the Fock space $\2H_0$, it is obvious
that $\al_0$ is extendible. Thus it remains to be proven the following:

\blem\label{lem53}
Each ultra local dynamics $\al^v\in\dyn(\6M)$ is extendible. 
\elem
\bpr
Let us consider any ultra local interaction $v$. 
For each test function \newline $f\in S(\7R)$, 
we introduce the unitary operator
\beqa
u(f|t):= e^{itv(f)}\otimes e^{itv(f)} \ \ \ .
\eeqa
Let $\2I$ be a bounded interval and denote by $\2I_\eps$, $\eps>0$, the 
enlarged interval \newline $\2I+(-\eps,\eps)$. 
We choose test functions $\chi^{(\2I,\eps)}\in S(\7R)$ such that 
\beqa
\chi^{(\2I,\eps)}(\1x)=\left\{
\begin{array}{ll}
1 & \1x\in \2I \\
0 & \1x\in \2I_\eps^c=\2I_\eps\bs\7R
\end{array} \right.  \  \ .
\eeqa
For an interval $\hat\2I\supset\2I_\eps$, the region 
$\hat\2I_\eps\bs\2I_\eps$ consists of two 
connected components $(\hat\2I_\eps\bs\2I_\eps)_\pm$ and there exist
test functions $\chi^\pm\in S(\7R)$ with 
\beqa
\begin{array}{l}
\supp(\chi^-)\subset (\hat\2I_\eps\bs\2I_\eps)_-\subset\2I_{LL}
\vs 
\supp(\chi^+)\subset (\hat\2I_\eps\bs\2I_\eps)_+\subset\2I_{RR}
\vs
\chi^{(\hat\2I,\eps)}-\chi^{(\2I,\eps)}=\chi^++\chi^- \ \ .
\end{array}
\eeqa
Let us write
\beqa
\mbox{$u(\2I,\eps|t):=u(\chi^{(\2I,\eps)}|t)$ 
\ and \ $u_\pm(t):=u(\chi^\pm|t)$.}
\eeqa
Since we have $[u(f_1|t),u(f_2|t)]=0$ for any pair of test functions
$f_1,f_2\in S(\7R)$, we obtain for each $\eps>0$ and for 
$\2I_\eps\subset\hat\2I$:
\beq
u(\hat\2I,\eps|t)= u(\2I,\eps|t)u_-(t)u_+(t)
\eeq
If we make use of the fact that $u_+(t)$ is 
$\al_F$-invariant and localized in $\2I_{RR}$, we conclude that 
$\te_\2I$ and $u_\pm(t)$ commute. Thus we 
obtain
\beq
\Ad(u(\hat\2I,\eps|t))\te_\2I= \Ad(u(\2I,\eps|t))\te_\2I
\eeq
which depends only of the localization interval $\2I$ since 
$\eps>0$ can be chosen arbitrarily small.
According to Lemma \ref{lem522}, the automorphisms 
\beqa
\hat\al_t^v:\hat\6M(\2I)\ni a\mapsto \Ad(u(\2I,\eps|t))a\in\hat\6M(\2I)
\eeqa
define a dynamics of $\hat\6M$ whose restriction to 
$\6M\olt\6M$ is $\al^v\otimes\al^v$. Thus $\al^v$ is extendible.
\epr

If $\hat\al_0$ denotes the natural extension of the free dynamics
$\al_0^{\otimes 2}$ to 
$\hat\6M$ and let $\hat \al^v$ be the extension of the 
ultra local dynamics $\al^v\otimes\al^v$ then, 
by using the Trotter product, we conclude that the dynamics 
\beqa
\hat\al:=\hat \al_0\times \hat \al^v
\eeqa
is an extension of the dynamics $(\al_0\times \al^v)^{\otimes 2}$ to 
$\hat\6M$. This leads to the following result:
\bpro\label{pro53}
Each dynamics of a $P(\phi)_2$-model is extendible.
\epro
\bpr
The statement follows from Lemma \ref{lem53} 
and from the fact that each dynamics 
of a $P(\phi)_2$-model is a Trotter product of the free dynamics $\al_0$
and an ultra local dynamics $\al_1$. 
\epr

The existence of interpolating kink states 
in $P(\phi)_2$-models is an immediate consequence of Proposition \ref{pro53}.

\bcor
Let $\al\in\dyn(\6M)$ be a dynamics of a $P(\phi)_2$-model. Then 
for each pair of vacuum states $\om_1,\om_2\in\6S_0(\al,\6M)$ there
exists an interpolating kink state $\om\in\6S(\om_1,\om_2)$.
\ecor
\bpr
By Proposition \ref{pro53} 
each dynamics of a $P(\phi)_2$-model is extendible and we can apply 
Corollary \ref{cor52} which implies the result.
\epr

\subsection{The Dynamics of the $\mbox{Yukawa}_2$ Model}
\label{s522a}

Since the dynamics of a $\mbox{Yukawa}_2$-like model 
can not be written as a Trotter product which consists of  
a free and an ultra local dynamics, it is a bit more 
complicated to show that these dynamics are extendible. 
We briefly summarize here the construction of the $\mbox{Yukawa}_2$ dynamics
which has been carried out by J. Glimm and A. Jaffe \cite{GlJa1}. 
We also refer to the work of R. Schrader \cite{Schra0,Schra1}. 

Let $\6M_s$ and $\6M_a$ be the 
nets of Cauchy data for the free Bose and Fermi field, represented 
on the Fock spaces $\2H_s$ and $\2H_a$ respectively. The Cauchy data 
of the $\mbox{Yukawa}_2$ model are given by the 
W*-tensor product $\6M:=\6M_s\olt\6M_a$ of the nets $\6M_s$ and $\6M_a$. 
Moreover, we set $\2H_0:=\2H_s\otimes\2H_a$.

\paragraph*{\it Step 1:}
In the first step, a Hamiltonian, which
is regularized by an UV-cutoff $c_0>0$ and an IR-cutoff $c_1>1$, $c_0<< c_1$,
is constructed.
For this purpose, one chooses test functions 
$\delta_{c_0},\chi_{c_1}\in S(\7R)$ with the properties:
\bdes
\itno a
\beqa
\supp(\delta_{c_0})\subset (-c_0,c_0) \ \ \mbox{ and } \ \ 
\int \8d\1x \ \delta_{c_0}(\1x)=1
\eeqa
\itno b
\beqa
\supp(\chi_{c_1})\subset (-c_1-1,c_1+1)  \ \ \mbox{ and } \ \ 
\chi_{c_1}(\1x)=1 
\\ 
\mbox{ for each $\1x\in (-c_1,c_1)$.}
\eeqa
\edes
The UV-regularized fields are given by
\beq\label{uvreg1}
\phi(c_0,\1x):=(\phi * \delta_{c_0})(\1x)
\ \ \mbox{ and } \ \ 
\psi(c_0,\1x):=(\psi * \delta_{c_0})(\1x)
\eeq
where $\phi$ is a massive free Bose field and 
$\psi$ a free Dirac spinor field at $t=0$. The fields, defined by 
equation (\ref{uvreg1}), act on $\2H_0$ via the operators
\beqa
\Phi(c_0,\1x):=\phi(c_0,\1x)\otimes\11_{\2H_a}
\ \ \mbox{ and } \ \ 
\Psi(c_0,\1x):=\11_{\2H_s}\otimes\psi(c_0,\1x) \ \ .
\eeqa

The regularized Hamiltonian $H(c_0,c_1)$ 
can be written as a sum of three parts: 
\bdes
\itno 1 
The free Hamiltonian $H_0$ which is given by 
\beqa
H_0=H_{0,s}\otimes\11_{\2H_a}+\11_{\2H_s}\otimes H_{0,a}
\eeqa
where $H_{0,s}$ and $H_{0,a}$ are the free Hamilton operators 
of the Bose and the Fermi field respectively.
\itno 2
The regularized Yukawa interaction term:
\beqa
H_Y(c_0,c_1)=\int \8d\1x \ \chi_{c_1}(\1x) \ \Phi(c_0,\1x)\ 
:\bar\Psi(c_0,\1x)\Psi(c_0,\1x):
\eeqa
\itno 3
The counterterms:  
\beqa
H_C(c_0,c_1)=
\sum_{n=0}^N \ z_n(c_0) \ \int \8d\1x \  \chi_{c_1}(\1x) 
\ :\Phi(\1x)^n:
\eeqa
where $z_n(c_0)$ are suitable renormalization constants.
\edes

The following statement has been established by J.Glimm and A.Jaffe 
\cite{GlJa1,GlJa3}:
\bthe
The counterterms $H_C(c_0,c_1)$ can be chosen in such a way that 
\bdes
\itno 1
the cutoff Hamiltonian $H(c_0,c_1)=(H_0+H_Y(c_0,c_1)+H_C(c_0,c_1))^{**}$
is a positive and self adjoint operator with domain $D(H_0)$.
\itno 2
The uniform limit 
\beqa
R(c_1,\zeta)=\lim_{c_0\to 0} (H(c_0,c_1)-\zeta)^{-1}
\eeqa
is the resolvent of a self adjoint operator $H(c_1)$.
\itno 3
$H(c_1)$ is the limit of $H(c_0,c_1)$ in the strong graph topology.
\edes 
\ethe

\bnoa
In the sequel, we shall use the following notation:
\beqa
u(c_0,c_1,t):=\exp(itH(c_0,c_1)) \ \ \mbox{ and } \ \ 
u(c_1,t):=\exp(itH(c_1)) \ \mbox{.}
\eeqa
\enoa

\brem
The aim is to show that $H(c_1)$ induces a dynamics of $\6M$, given locally 
by the equation 
\beqa
\al_t|_{\6M(\2I)}=\Ad( \ u(c_1,t) \ ) \ \ \ 
\mbox{ for $\2I_{|t|}:=\2I+(-|t|,|t|)\subset (-c_1,c_1)$.}
\eeqa
However, in comparison to the $P(\phi)_2$-models, 
there are some more technical difficulties which have to be overcome.
\bdes
\itno {i} 
The Hamiltonian $H(c_1)$ is only defined as a limit of the 
Hamiltonians $H(c_0,c_1)$ and it has no mathematical meaning when 
written as a sum
\beqa
H_0+H_Y(c_1)+H_C(c_1) \ \ \ .
\eeqa
Thus the construction scheme for a dynamics, 
as it has been used for $P(\phi)_2$-models, 
does not apply.
\itno {ii}
On the other hand, one might try to apply  $P(\phi)_2$-like methods to the 
Hamiltonian $H(c_0,c_1)$, for which the UV-cutoff is not removed.
For this purpose, one wishes to write $H(c_0,c_1)$ as a sum \newline
$H(c_0,c_1)=H_1(c_0,c_1)+H_2(c_0,c_1)$ where  
$H_1(c_0,c_1)$ induces a dynamics 
$\al_1$ with propagation speed $\ps(\al_1)\leq 1$ and 
$H_2(c_0,c_1)$ induces a dynamics $\al_2$ with propagation 
speed $\ps(\al_2)=0$.

The difficulty with writing such a decomposition for $H(c_0,c_1)$ arises 
from the fact that the Yukawa interaction term $H_Y(c_0,c_1)$ induces 
an automorphism group with infinite propagation speed. 
\edes
\erem 

\paragraph*{\it Step 2:} 
In the next step, one introduces test functions 
$\chi_{(\2I,s,c_0)}$, 
depending on a bounded interval $\2I$, a real number $s>0$ 
and the UV-cutoff $c_0$, fulfilling the conditions 
\beq\label{cutoffreg}
\begin{array}{l}
\supp(\chi_{(\2I,s,c_0)})\subset \2I_{2c_0+|s|+\epsilon}\bs\2I_{|s|-\epsilon}
\ \ \mbox{ and } 
\vs\vs \ \ 
\chi_{(\2I,s,c_0)}(\1x)=1 \ \ 
\mbox{ if $\1x\in\2I_{2c_0+|s|}\bs\2I_{|s|}$.}
\end{array}
\eeq
Here $\epsilon<< c_0$ is any sufficiently small positive number. 
The Hamiltonian $H(c_0,c_1)$ is replaced by the operator 
\beq
H(\2I,s,c_0,c_1):=H_0+H_C(c_0,c_1)+H_Y(\2I,s,c_0,c_1)
\eeq
depending additionally on  
$\2I$ and $s$, where $H_Y(\2I,s,c_0,c_1)$ is given by
\beqa
H_Y(\2I,s,c_0,c_1):=\int \8d\1x \ 
\Phi(c_0,\1x)\ :\bar\Psi(c_0,\1x)\Psi(c_0,\1x): \ 
( \ \chi_{c_1}(\1x)-\chi_{(\2I,s,c_0)}(\1x) \ ) \ \ \ .
\eeqa

In order to 
construct from these data a $c_1$-independent approximation of 
the dynamics which maps 
$\6M(\2I)$ onto $\6M(\2I_{|t|})$, one defines the unitary operators
\beqa
w(\2I,c_0,c_1,t):=\prod_{j=1}^n
\exp\biggl ( i{t\over n} H( \ \2I,(n-j)n^{-1}t,c_0,c_1 \ ) \biggr )
\eeqa
where $n$ is equal to the integral part of $|c_0^{-1}t|$. The  
lemma, given below, has been established in \cite{GlJa1}. 

\blem\label{lem59}
{\em\cite[Lemma 9.1.2]{GlJa1}}
The adjoint action of $w(\2I,c_0,c_1,t)$ induces an automorphism 
\beqa
\al_t^{(\2I,c_0)}:=\Ad(w(\2I,c_0,c_1,t)):\6M(\2I)\to\6M(\2I_{|t|})
\eeqa
which is independent of $c_1$.
\elem

\paragraph*{\it Step 3:}
For technical reasons, to control convergence as 
$c_0$ tends to zero, the length of time propagation is scaled, and one defines
for $\lam\in [0,1]$ the $c_1$-independent automorphism
\beqa
\al^{(\2I,c_0,\lam)}_t:=\Ad(w(\2I,c_0,c_1,\lam,t)):\6M(\2I)\to\6M(\2I_{|t|})
\eeqa
where $w(\2I,c_0,c_1,\lam,t)$ is given by
\beqa
w(\2I,c_0,c_1,\lam,t):=\prod_{j=1}^n
\exp\biggl ( i{\lam\cdot t\over n} H(\2I,(n-j)n^{-1}t,c_0,c_1) \biggr ) \ \ .
\eeqa
The final approximation is given by averaging over $\lam$:
\beqa
\al^{(\2I,c_0,\ell)}_t(a):=\int \8d\lam \ f_{\ell}(\lam) \ 
\al^{(\2I,c_0,\lam)}_t(a)
\eeqa
where $f_\ell$ is a positive continuous function such that 
\beqa
\int \8d\lam \ f_\ell(\lam) =1 \ \ 
\mbox{ and $\supp(f_\ell)\subset [1-\ell,1]$, $\ell\leq 1$.}  
\eeqa

Finally, J. Glimm and A. Jaffe have established the result: 
\bthe\label{theGlJa913}{\em \cite[Theorem 9.1.3]{GlJa1}}
There exists a function $c:\ell\mapsto c_\ell $ with 
$\lim_{\ell\to 0}c_\ell =0$ such that  
\beq
\al^Y_t(a):=w-\lim_{\ell\to 0}\al^{(\2I,c_\ell ,\ell)}_t(a)
=u(c_1,t) \ a \ u(c_1,t)^*
\eeq
for each $a\in\6M(\2I)$ and for each sufficiently large $c_1$.
\ethe

\subsection{Kink States in Models with $\mbox{Yukawa}_2$ Interaction}
\label{s522}

We shall use an analogous strategy as above (step 1- step 3) in order to show 
that the dynamics $\al^Y$, which is given due to 
Theorem \ref{theGlJa913} is extendible.

\bthe\label{the55}
The dynamics $\al^Y$ of the $\mbox{Yukawa}_2$ model is 
extendible.
\ethe

Let us prepare the proof. First, we give a few comments on 
the notation to be used.

\bnoa
\bdes
\itno {a}
In the sequel, we write $\hat w(\cdots)=w(\cdots)^{\otimes 2}$ 
and $\hat u(\cdots)=u(\cdots)^{\otimes 2}$ for the corresponding quantities 
of the two-fold theory. 
As in step 3 above, we also define the automorphism 
\beqa
\hat\al^{(\2I,c_0,\lam)}_t:=\Ad(\hat w(\2I,c_0,c_1,\lam,t))
\eeqa
and the average
\beqa
\hat\al^{(\2I,c_0,\ell)}_t(a)=\int \8d\lam \ f_\ell(\lam) \ 
\hat\al^{(\2I,c_0,\lam)}_t(a) \ \ \ .
\eeqa
\itno {b}
Let $\om_0$ be the vacuum state with respect to the free dynamics which is 
induced by $H_0$. We denote by 
$\Psi_\2I$ the universal localizing map 
of the standard split inclusion $\Lam(\2I)\otimes\Lam(\2I)$ 
and we define $\te_\2I:=\Psi_\2I(\11\otimes u_F)$.
\edes
\enoa
 
\blem\label{lem510}
The adjoint action of 
$\hat w(\2I,c_0,c_1,t)$ induces an automorphism 
\beqa
\hat\al_t^{(\2I,c_0)}:
\hat\6M(\2I)\to \hat\6M(\2I_{|t|})
\eeqa
which is independent of $c_1$. 
\elem
\bpr
By Lemma \ref{lem59}, it is sufficient to prove that 
\beqa
\Ad(\hat w(\2I,c_0,c_1,t))\te_\2I 
\eeqa
is $c_1$-independent. 
Indeed, following the arguments in the proof of 
Proposition \ref{pro53}, we conclude that 
\beqa
\te'_\2I:=\exp(i\tau H(\2I,s,c_0,c_1))^{\otimes 2} \ \te_\2I \ 
\exp(-i\tau H(\2I,s,c_0,c_1))^{\otimes 2}
\eeqa
is $c_1$-independent for $|\tau|\leq c_0$ and that $\te'_\2I$ is contained in 
$\hat\6M(\2I_{|s|+|\tau|})$.
Composing $n$ such maps, we obtain the lemma. 
\epr

In complete analogy to Theorem \ref{theGlJa913} we have:
\blem\label{lem511}
\beqa
\hat\al^Y(a):= w-\lim_{\ell\to 0}\hat\al^{(\2I,c_\ell ,\ell)}_t(a)=
\hat u(c_1,t) \ a \  \hat u(c_1,t)^*  
\eeqa
For each $a\in \hat\6M(\2I)$ and for each 
sufficiently large $c_1$.
\elem
\bpr
By Theorem \ref{theGlJa913}, we conclude that the lemma holds for each 
$a\in\6M(\2I)\olt\6M(\2I)$. Hence it remains to be proven that 
\beqa
w-\lim_{\ell\to 0}\hat\al^{(\2I,c_\ell ,\ell)}_t(\te_\2I)=
\hat u(c_1,t) \ \te_\2I \ \hat u(c_1,t)^* \ \mbox{ .} 
\eeqa
The Corollary 9.1.9 of \cite{GlJa1} states:
\beqa
w-\lim_{\ell\to 0} \int \8d\lam \ 
( \ \hat w(\2I,c_\ell ,c_1,\lam,t) \ - \ \hat u(c_\ell ,c_1,\lam t) \ )
f_\ell(\lam)=0 \ \ .
\eeqa
We define 
\beqa
\te_\2I(\ell,t):=\hat\al^{(\2I,c_\ell ,\ell)}_t(\te_\2I)
\ \ \mbox{ and } \ \ 
\bar\te_\2I(\ell,t):=\int \8d\lam \ f_\ell(\lam) \ 
\Ad( \ \hat u(c_\ell ,c_1,\lam t) \ )\te_\2I \ \ \ .
\eeqa
The Schwarz's inequality implies for each $\psi\in\2H_0\otimes\2H_0$:
\beqa\begin{array}{l}
|\<\psi,\te_\2I(\ell,t)-\bar\te_\2I(\ell,t)\psi\>|
\vs
\leq
2||\psi||\cdot\biggl (\int \8d\lam \ f_\ell(\lam) \ 
||( \ \hat w(\2I,c_\ell ,c_1,\lam,t) \ 
- \ \hat u(c_\ell ,c_1,\lam t) \ )\psi||^2\biggr )^{1/2}
\end{array}\eeqa
Since $||(v-u)\psi||^2=2\cdot \8{Re}(\<(v-u)\psi,u\psi\>)$, we obtain:
\beqa
|\<\psi,\te_\2I(\ell,t)-\bar\te_\2I(\ell,t)\psi\>| \ \ \leq
\ \ 4||\psi||\cdot\biggl (\int \8d\lam \ f_\ell(\lam) \ 
\vs\vs
\8{Re}\biggl\<\ (\hat w(\2I,c_\ell ,c_1,\lam,t) \ 
- \ \hat u(c_\ell ,c_1,\lam t) \ )\psi , 
\hat u(c_\ell ,c_1,\lam t)\psi \biggr\> \biggr )^{1/2}
\eeqa
which proves the lemma.
\epr

\paragraph*{\it Proof of Theorem \ref{the55}:}
We conclude from Lemma \ref{lem511} and Lemma \ref{lem522}
that the automorphism group $\hat\al^Y$ is a dynamics of the extended 
net $\hat\6M$ whose restriction to 
$\6M\olt\6M$ is $\al^Y\otimes\al^Y$. Thus $\al^Y$ is extendible.
\epr

\brem
According to \cite{Schra1}, each dynamics $\al^{Y+P}$ 
of a quantum field theory model with 
$\mbox{Yukawa}_2$ plus $P(\phi)_2$ boson self-interaction is extendible.
\erem

Finally, we conclude from Theorem \ref{the55}:

\bcor\label{corll55}
Let $\al^{Y+P}$ be a dynamics of a quantum field theory model with 
$\mbox{Yukawa}_2$ plus $P(\phi)_2$ boson self-interaction.
For each pair $\om_1,\om_2$ of vacuum states with respect to $\al^{Y+P}$,
there exists a kink state $\om$ in $\6S(\al^{Y+P}|\om_1,\om_2)$.
\ecor
\subsection{Wess-Zumino Models}
\label{s523a}

One interesting class of quantum field theory models which possess 
more than one vacuum sector are the $N=2$ Wess-Zumino models in 
two-dimensional space-time. 
Their properties have been studied 
in several papers \cite{JanWeit0,JaLesWeit1,JaLesWeit2,JanWeit1,JaLes} 
and  we summarize the main results which are 
established there in order to setup our subsequent analysis. 

The field content of these models with a finite volume cutoff $c>0$
consists of one {\em complex}
Bose field $\phi_c$ and one Dirac spinor field $\psi_c$, acting 
as operator valued distributions on the Fock spaces 
\beqa
\2H_a(c) &:=& \bigoplus_{n=0}^\infty \ L_2(T_c,\7C^2)^{\otimes_a}
\vs\vs
\2H_s(c) &:=& \bigoplus_{n=0}^\infty \ L_2(T_c,\7C)^{\otimes_s}
\eeqa
where $a,s$ stands for 
symmetrization or anti-symmetrization of the tensor product and 
$L_2(T_c,\7C^k)$ ($k=1,2$) denotes the Hilbert space of $\7C^k$-valued and 
square integrable functions, living on the 
circle $T_c$ of length $c$. The net of Cauchy data for the finite 
volume theory is given by 
\beqa
\6M_c:(-c,c)\supset\2I \mapsto \6M_c(\2I)=\6M_{c,s}(\2I)\olt\6M_{c,a}(\2I)
\eeqa
where the nets $\6M_{c,s}$ and $\6M_{c,a}$ are defined by the assignments: 
\beqa
\6M_{c,s}:(-c,c)\supset\2I\mapsto \6M_{c,s}(\2I):=
\biggl\{ e^{i(\phi_c(f_1)+\pi_c(f_2))}\biggm | \supp (f_j)\subset \2I\biggr\}''
\eeqa
\beqa
\6M_{c,a}:(-c,c)\supset\2I\mapsto \6M_{c,a}(\2I):=
\biggl\{ \psi_c(f_1),\bar\psi_c(f_2)
\biggm | \supp (f_j)\subset \2I\biggr\}'' 
\eeqa
where $\pi_c$ is the canonically conjugate of $\phi_c$.

Let $\6M:=\6M_{c=\infty}$ be the net of Cauchy data in the infinite 
volume limit, then the map 
\beqa
\iota_c:\bmat\phi(f_{11}) & \pi(f_{12}) \\
      \psi(f_{21}) & \bar\psi(f_{22}) \\ \emat \mapsto
\bmat\phi_c(f_{11}) & \pi_c(f_{12}) \\
      \psi_c(f_{21}) & \bar\psi_c(f_{22}) \\ \emat
\ \ ; \ \ \supp(f_{ij})\subset (-c,c)
\eeqa
is a W*-isomorphism 
which identifies the nets $\6M$ and $\6M_c$ for those regions $\2I$ which are  
contained in $(-c,c)$. 

The interaction part of the {\em formal} Hamiltonian consists 
of two parts.
\bdes
\itno a
A $P(\phi)_2$-like part:
\beqa
H_P(v,c)=\int_{T_c} \8d \1x \ :|v'(\Phi_c)|^2: \ - \ :|\Phi_c|^2:
\eeqa
\itno b
A $\mbox{Yukawa}_2$-like part:
\beqa
H_Y(v,c):=\int_{T_c} \8d \1x \ : \bar\Psi_c
\bmat 
v''(\Phi_c)-1 & 0 \\
0 & v''(\Phi_c)^*-1 \\
\emat
\Psi_c :
\eeqa
\edes
where $v$ is a polynomial of degree $\deg(v)=n$, 
called {\em superpotential}, and 
the fields $\Phi_c$ and $\Psi_c$ are given by 
\beqa
\Phi_c:=\phi_c\otimes\11_{\2H_a(c)} \ \ \mbox{ and } \ \ 
\Psi_c:=\11_{\2H_s(c)}\otimes\psi_c \ \ . 
\eeqa

According to the results of \cite{JanWeit0,JaLes,JaLesWeit1,JaLesWeit2}, 
it has been shown that, 
there is a self-adjoint Fredholm operator $Q(v,c)$, called 
{\em supersymmetry generator}, on \newline 
$\2H_0(c):=\2H_s(c)\otimes\2H_a(c)$.
The {\em Fredholm index} of $Q(v,c)$
\beqa
\ind(Q(v,c))=\dimker(Q(v,c))-\dimcoker(Q(v,c)) 
\eeqa
has been computed in \cite{JaLesWeit1}. The result is
\beqa
|\ind(Q(v,c))|=\deg(v)-1 \ \ \ .
\eeqa
The space $\2H_0(c)$ may be decomposed $\2H_0(c)=\2H_+(c)\oplus\2H_-(c)$ into
the eigenspaces of the fermion parity operator $\Gam:=(-1)^{N_a}$, where 
$N_a$ is the fermion number operator. 
With respect to this decomposition, the operator $Q(v,c)$ has the form 
\beqa
Q(v,c)=\bmat 0 & Q_+(v,c) \\ Q_-(v,c) & 0 \\ \emat \ \ . 
\eeqa
The full Hamiltonian of the finite volume model is given by 
\beqa
H(v,c)=Q(v,c)^2
\eeqa
which implies:
\beqa
\dimker(H(v,c))=|\dimker(Q_+(v,c))-\dimker(Q_-(v,c))|=\deg(v)-1
\eeqa

The Hamiltonian $H(v,c)$ induces a dynamics $\al^{(v,c)}$ of the finite 
volume model and we conclude from the results of \cite{JanWeit0}:

\bthe\label{theJanWeit0}{\em\cite[Theorem 1]{JanWeit0}}
There exists at least $\deg(v)-1$ vacuum sectors with respect to the dynamics 
$\al^v:=\al^{(v,c=\infty)}$ of the model in the infinite volume limit. 
\ethe

\subsection{Kink States in Wess-Zumino Models}
\label{s523}

In order to prove the existence of kink sectors, we now apply 
the results which have been established in Section \ref{s521} and 
Section \ref{s522} to $N=2$ Wess-Zumino Models. 

\paragraph*{\it The case $\deg(v)=3$:}
Let us have a closer look at the simplest non-trivial case $\deg(v)=3$.
We let 
\beqa
v'(z)=\lam_2 z^2+\lam_1z+\lam_0 \ \ \ .
\eeqa
As in the previous sections (equation (\ref{uvreg1})), we introduce the 
UV-regularized fields:
\beqa
\Phi(c_0,\1x):=(\Phi * \delta_{c_0})(\1x)
\ \ \mbox{ and } \ \ 
\Psi(c_0,\1x):=(\Psi * \delta_{c_0})(\1x)
\eeqa
where $\delta_{c_0}$ is a smooth test function with support in $(-c_0,c_0)$.
We obtain for the $P(\phi)_2$-like part of the 
regularized interaction Hamiltonian
\beqa
\begin{array}{l}
H_P(v;c_0,c_1)=\int \8d \1x \chi_{c_1}(\1x) \biggl [
\vs\vs
:|\lam_2\Phi(c_0,\1x)^2+\lam_1\Phi(c_0,\1x)+\lam_0|^2:-:|\Phi(c_0,\1x)|^2: 
\biggr]
\end{array}
\eeqa
and for the $\mbox{Yukawa}_2$-like part:
\beqa
\begin{array}{l}
H_Y(v;c_0,c_1)=\int \8d \1x \chi_{c_1}(\1x) \biggl [  
\vs\vs
:\bar\Psi(c_0,\1x) \bmat 2\lam_2\Phi(c_0,\1x)+\lam_1-1 & 0 \\
0 & 2\bar\lam_2\Phi(c_0,\1x)^*+\bar\lam_1-1 \\
\emat \Psi(c_0,\1x): \biggr ]
\end{array}
\eeqa

Using the same techniques as in Section \ref{s521} and 
Section \ref{s522}, we obtain the corollary
(see also Corollary \ref{corll55}): 

\bcor
Let $v$ be a superpotential of degree $\deg(v)=3$. Then the following 
statements are true:
\bdes
\itno 1
The dynamics $\al^v\in\dyn(\6M)$  
of the model in the infinite volume limit
is extendible.
\itno 2
There exists two different vacuum sectors
$e_1,e_2\in\sec_0(\al^v,\6M)$ and two different
kink sectors $\te\in\sec(e_1,e_2)$, $\bar\te\in\sec(e_2,e_1)$.
\edes
\ecor

\paragraph*{\it The case $\deg(v)>3$:}
We close this section by discussing the remaining case. 

In order to show the extendibility of $\al^v\in\dyn(\6M)$,
we can try to proceed in the same manner as for the case $\deg(v)=3$. 
According to Section \ref{s522} ({\it Step 2} and {\it Step 3}),
we construct an approximation 
\beqa
\6M(\2I)\olt\6M(\2I)\ni a\mapsto\hat\al^{(v;\2I,c_0,\ell)}_t(a)
:=\int \8d\lam \ f_{\ell}(\lam) \ \hat\al^{(v;\2I,c_0,\lam)}_t(a)
\eeqa
of the dynamics $\al^v\otimes\al^v$ of the two-fold theory.
Provided that the corresponding result of Lemma \ref{lem510} is true 
for the case $\deg(v)>3$ also, the linear maps 
$\hat\al^{(v;\2I,c_0,\ell)}_t$ can be extended to the algebra 
$\hat\6M(\2I)$. 

For the generalization of 
Theorem \ref{the55},
it seems that the most difficult part is to show that there exists a function 
$c:\ell\mapsto c_\ell $ with $\lim_{\ell\to 0}c_\ell =0$ such that  
\beq\label{dynapp}
\al^v_t(a):=w-\lim_{\ell\to 0}\al^{(v;\2I,c_\ell ,\ell)}_t(a)  \ \ .
\eeq
The regularized Yukawa-like part of the Hamilton density 
contains terms of the form
\beqa
&&:\Psi^{(i)}(c_0,\1x)\Psi^{(j)}(c_0,\1x): \ \ :\Phi(c_0,\1x)^k: 
\vs\vs 
&&\mbox{ $i,j\in\{0,1\}$, $i\not=j$ and $k\leq \deg(v)-2$, }   
\eeqa
where $\Psi^{(j)}$ denotes the $j$-component of the Dirac spinor 
field $\Psi$. Since there are contributions with $k>1$,
the proof of Theorem \ref{theGlJa913} does not directly apply.

Provided that for each superpotential $v$ the dynamics $\al^v$ is 
extendible, we conclude that
for each pair of vacuum sectors $e_1,e_2\in \sec_0(\al^v,\6M)$ there exists 
a kink state $\om\in\6S(e_1,e_2)$. Then the model 
possesses at least $\deg(v)(\deg(v)-1)$ different non-trivial 
kink sectors. 

\section{Conclusion and Outlook}
\label{s8}

In the present paper, a construction scheme for 
kink sectors has been developed which can be applied to a large 
class of quantum field theory models.
Most of the techniques which are used, except those in the proof of the 
extendibility of the dynamics, concern operator algebraic methods. 
They are model independent in the sense that they can be derived 
from first principles. 
There are still some interesting open problems and  
we shall make a few remarks on them here.  

\paragraph{Some further remarks on kink states:}
Let us consider a quantum field theory model ($P(\phi)_2$, $Y_2$),  
possessing vacua $\om_1,\om_2$ which are related by a symmetry $\chi$.
According to Theorem \ref{the51}, there exists an 
automorphism $\chi^\2I$ which induces a kink state 
$\om=\om_1\circ\chi^\2I$. Note that $\om$ is a pure state 
in this case. 

Alternatively, we obtain a kink state $\hat\om$ by passing to the 
two-fold tensor product of the theory with itself first and then 
by restricting the $\al_F$-interpolating 
automorphism $\be^\2I$, whose existence follows also from Theorem \ref{the51},
to the first tensor factor, i.e.: 
\beqa
\hat\om=\om_1\otimes\om_2\circ\be_{C^*(\6A)\otimes\11} \ \ .
\eeqa

Provided the split property for wedge algebras holds for 
the interacting vacua \cite{Schl4}, then, by applying a recent result 
of M. M\"uger \cite{Mue2}, we conclude that $[\hat\om]$ is 
nothing else but the infinite multiple of $[\om]$. 
  
\paragraph{The problem of reducibility:}
The problem of reducibility arises if the 
vacua under consideration are not related by a symmetry since then 
our construction scheme leads to kink representations of the form
\beqa
\pi=\pi_1\otimes\pi_2\circ\be|_{C^*(\6A)\otimes\11}
\eeqa
where $\be$ is an automorphism and $\pi_1,\pi_2$ are vacuum representations. 
The representation $\pi$ is not irreducible and whether $\pi$ can be 
decomposed into irreducible sub-representations is still an open 
problem. Some of our results \cite[Theorem 4.4.3]{Schl5} suggest 
that $\pi$ is, in non exceptional cases, an infinite multiple of one 
irreducible component. 

\paragraph{Kink sectors in $d>1+1$ dimensions:}
It would be desirable to apply our program 
to quantum field theories in higher dimensions. 
Let us suppose a theory, given by a net of W*-algebras $\6A$, 
possesses two locally normal vacuum states $\om_1,\om_2$.
 
As a sensible generalization of a kink states 
to $d>1+1$, we propose to consider  
locally normal states $\om$ which fulfill the   
interpolation condition:
\beq\label{intergen}
\om|_{C^*(\6A,S_1)}=\om_1|_{C^*(\6A,S_1)} \ \ \mbox{ and } \ \ 
\om|_{C^*(\6A,S_2')}=\om_2|_{C^*(\6A,S_2')}
\eeq
where $S_1,S_2$, $S_1\subset S_2$, are space-like cones. 
The state $\om$ describes the coexistence of two phases 
which are separated by the {\em phase boundary}
$\pa S:=S_1'\cap S_2$.

Let us assume duality for space-like cones in the vacuum representations 
under consideration. Furthermore, we presume that the inclusion 
\beqa
\Lam=(\6A_{\pi_1}(S_1),\6A_{\pi_1}(S_2),\Om_1)
\eeqa
is standard split. Here $(\2H_1,\pi_1,\Om_1)$ is a 
GNS-triple with respect to $\om_1$.

Unfortunately, for $d>1+1$ the phase boundary $\pa S$ is not compact 
and therefore our construction scheme can not directly 
be generalized to higher dimensions.

In order to overcome this difficulties, we consider a sequence of 
standard split inclusions 
\beqa
\Lam_n:=(\6A_{\pi_1}(\2O_{1n}),\6A_{\pi_1}(\2O_{2n}),\Om_1) 
\eeqa
where $\2O_{1n}\subset\subset \2O_{2n}$ are bounded double cones 
such that $\2O_{jn}$ tends to $S_j$ for $n\to\infty$. 

As in the $1+1$-dimensional case we pass now to the two-fold 
tensor product of the theory with itself. Denote by 
$\Psi_{\Lam_n\otimes\Lam_n}$ the universal localizing map 
with respect to the inclusion $\Lam_n\otimes\Lam_n$. 
Since the operators 
\beqa
\te_n:=\Psi_{\Lam_n\otimes\Lam_n}(\11\otimes u_F)
\eeqa
are localized in a bounded region, we may define the following 
automorphisms of $C^*(\6A)$: 
\beqa
\be_n:=(\pi_1\otimes\pi_1)^{-1}\circ\be_n\circ(\pi_1\otimes\pi_1) \ \ .
\eeqa
We obtain a sequence of states $\{\om_n,n\in\7N\}$ where $\om_n$ 
is given by:
\beqa
\om_n:=\om_1\otimes\om_2\circ\be_n|_{C^*(\6A)\otimes\11} \ \ .
\eeqa
For large $n$ the states $\om_n$ have {\em almost} the correct interpolation 
property, namely for each pair of local 
observables $a,b$ where $a$ is localized $S_2'$ and $b$ is localized in $S_1$,
there exists a sufficiently large $N$ such that
\beqa
\om_n(a)=\om_1(a) \ \ \mbox{ and } \ \ \om_n(b)=\om_2(b) 
\eeqa
for each $n>N$. 
Note that each state $\om_n$ fulfills the 
Borchers criterion since $\om_n$ belongs to the vacuum sector $[\om_1]$.
 
In order to obtain generalized kink states, we propose to investigate  
weak*-limit points of the sequence $\{\om_n,n\in\7N\}$. 
Note that each weak*-limit $\om_\iota$ point
of the sequence $\{\om_n,n\in\7N\}$ fulfills the interpolation 
condition (\ref{intergen}). 
It remains to be proven that the weak*-limit points are locally normal.


\subsubsection*{{\it Acknowledgment:}}
I am grateful to Prof. K. Fredenhagen for 
supporting this investigation with many ideas.
I am also grateful to Dr. K.H. Rehren for many hints and discussion.
This investigation is financially supported by the Deutsche 
Forschungsgemeinschaft who is also gratefully acknowledged.



\end{document}